\documentclass{emulateapj}

\newcommand{\msun}{M_{\odot}}

\shorttitle{Cluster complexes at $z~1.5$}
\shortauthors{Adamo et al.}

\begin{document}


\title{High resolution study of the cluster complexes in a lensed spiral at  redshift ~1.5; \\
constraints on the bulge formation and disk evolution\altaffilmark{1}}


\author{Angela Adamo\altaffilmark{2}, G. \"Ostlin\altaffilmark{3}, N. Bastian\altaffilmark{4}, E. Zackrisson\altaffilmark{3}, R. C. Livermore\altaffilmark{5}, L. Guaita\altaffilmark{3}}
\email{adamo@mpia.de}
\altaffiltext{1}{Based on observations made with the NASA/ESA Hubble Space Telescope, obtained at the Space Telescope Science Institute, which is operated by the Association of Universities for Research in Astronomy, Inc., under NASA contract NAS 5-26555. These observations are associated with program 12065}
\altaffiltext{2}{Max-Planck-Institut for  Astronomy, K\"onigstuhl 17, D-69117 Heidelberg, Germany}
\altaffiltext{3}{Department of Astronomy, Stockholm University, SE-10691 Stockholm, Sweden}
\altaffiltext{4}{Astrophysics Research Institute, Liverpool John Moores University, Egerton Wharf, Birkenhead, CH41 1LD, UK}
\altaffiltext{5}{Institute for Computational Cosmology, Durham University, South Road, Durham DH1 3LE, UK}




\begin{abstract}
We analyse the clump population of the spiral galaxy Sp\,1149 at redshift 1.5. Located behind the galaxy cluster MACS J1149.5+2223, Sp\,1149 has been significantly magnified allowing us to study the galaxy on physical scales down to $\sim100$ pc. The galaxy cluster frame is among the targets of the Cluster Lensing And Supernova survey with Hubble ({\it CLASH}), an ongoing {\it Hubble Space Telescope} (HST) Multi-Cycle Treasury program. We have used the publicly available multi-band imaging dataset to reconstruct the spectral energy distributions (SEDs) of the clumps in Sp\,1149, and derive, by means of stellar evolutionary models, their physical properties. We found that 40 \% of the clumps observed in Sp\,1149 are older than 30 Myr and can be as old as 300 Myr. These are also the more massive (luminous) clumps in the galaxy. Among the complexes in the local reference sample, the star-forming knots in luminous blue compact galaxies could be considered progenitor analogs of these long-lived clumps. The remaining 60 \% of clumps have colors comparable to local cluster complexes, suggesting a similar young age. We observe that the Sp\,1149 clumps follow the $M\propto R^2$ relation similar to local cluster complexes, suggesting similar formation mechanisms although they may have different initial conditions (e.g. higher gas surface densities). We suggest that the galaxy is experiencing a slow decline in star formation rate and a likely transitional phase toward a more quiescent star-formation mode. The older clumps have survived between 6 and 20 dynamical times and are all located at projected distances smaller than 4 kpc from the centre. Their current location suggests migration toward the centre and the possibility to be the building blocks of the bulge. On the other hand, the dynamical timescale of the younger clumps are significantly shorter, meaning that they are quite close to their birthplace. We show that the clumps of Sp 1149 may account for the expected metal-rich globular cluster population usually associated with the bulge and thick disk components of local spirals.

\end{abstract}


\section{Introduction}

Recent results suggest that galaxies which are still star-forming at the present time (redshift, $z\sim0$), e.g. late-type galaxies such as spirals have assembled more than 80 \% of their stellar mass between $0\lesssim z\lesssim2$ \citep[e.g.][]{2012MNRAS.tmp..321M, 2012ApJ...752...41Y, 2012ApJ...745..149L}. It is well known the the cosmic star formation rate density shows a peak at $z\sim2$ and a clear decline between $0\lesssim z\lesssim2$ \citep[][for one of the updated diagrams of the cosmic star formation rate as function of $z$]{1996ApJ...460L...1L, 1996MNRAS.283.1388M, 2011ApJ...737...90B}. More recently it has become clear that star formation at all $z$ is probably dominated by normal, $L^*$ star forming galaxies \citep{2007ApJ...660L..43N, 2011ApJ...742...96W}. From those two results it follows that normal spiral galaxies have strongly declining SFRs between $z=2$ and the present.



At high $z$, galaxy morphologies do not follow the classical Hubble classification. Galaxies are intrinsically different and show features, like very massive star-forming clumps that hardly could be associated to any star-forming region we observe in local galaxies. Using HST ultra deep field (UDF) data, \citet{2005ApJ...631...85E} defined six morphological types, of which four (chain galaxies, clump clusters, double, tadpole) are based on the prominent presence of compact clumps, and absence of any bulge or exponential profile. In particular, chain and clump cluster galaxies are probably the same class of objects but observed at different inclinations. At $z\geq 1$, clumps are also observed in galaxies which show structural and kinematics properties typical of disks \citep[e.g.][]{2007ApJ...658..763E, 2010MNRAS.404.1247J, 2011ApJ...733..101G}. It seems, therefore, that clump galaxies, in the absence of major mergers, will become  the high $z$ progenitors of modern disk galaxies \citep[e.g.][]{2009ApJ...701..306E}. High-$z$ clumpy disks are observed to have high gas fractions, with spatially extended distributions that overlap the UV morphologies, and clear hints of rotation \citep[e.g.][]{2010ApJ...713..686D, 2010Natur.463..781T}. Observationally, there are hints that the \citet{1964ApJ...139.1217T} $Q$-parameter of clumpy disks is $\lesssim 1$, suggesting that clumps form because of gravitational instability \citep[e.g.][]{2010MNRAS.404.1247J, 2011ApJ...733..101G}.  At the resolution achieved by these studies, the masses of the clumps are quite high ($10^7-10^9 \msun$), with the radii between 0.5 and a few kpc. Theoretical models and simulations predict that cold accretion of gas could at the same time feeds the galaxy prolonging the gas depletion time, assuming that the star formation rate (SFR) stays constant, and induces instabilities in the gas to form clumps \citep[see][for a short review]{2011EAS....51...59E}. In this scenario, the average surface density of SFR becomes a direct measure of the gas surface density present in the galaxy \citep{2012arXiv1209.5741L}. Under the assumption of a rotating and marginally unstable ($Q\sim1$) disk, \citet{2012arXiv1209.5741L} argued that the mass of the more massive clumps which can form increases for higher disk surface density. This finding can, indeed, explain why when the properties of high $z$ clumps are compared to local star-forming regions, they appear to follow the same physical relations as the latter but at much higher scales of SFR, luminosity, total stellar mass, etc \citep[e.g.][]{2010Natur.464..733S, 2010Natur.463..781T}.

A comparison study of the clump properties in clump clusters, chains and young spirals shows that the former may evolve into spiral galaxies \citep{2009ApJ...692...12E}. Indeed, theoretical works and simulations show that clumps form in gravitationally unstable, highly turbulent disks \citep[e.g.][]{2008A&A...486..741B, 2010MNRAS.404.2151C}. These clumps are expected to migrate toward the centre of the system because of dynamical friction and tidal torques in about 10 times the dynamical time, i.e. a total time range of  0.5--1 Gyr. In the centre clumps  coalesce and slowly build up the bulge and the thick disk component \citep{2007ApJ...670..237B, 2008ApJ...688...67E}. However, this scenario works if clumps survive long enough to reach the centre of the galaxy. In simulations and analytical studies, different initial conditions and assumptions lead to opposite fates of the clumps, i.e. they can be rapidly disrupted by stellar feedback \citep{2010ApJ...709..191M}, or survive to the initial gas loss for several dynamical times, so they could spiral inward \citep{2010MNRAS.406..112K}. Streams of material coming from dissolving clumps could still migrate toward the centre, but the bulge formation mechanism would be more inefficient \citep{2011ApJ...733..101G}.  

In spite of the progress made in understanding how galaxies have evolved from the time at which the cosmic SFR peaked ($z \sim 2$) to the current decline, it is still challenging to get a complete census of the whole process.  The major obstacles are the limited resolution and sensitivity, which create a bias toward galaxies with extreme SFRs and physical conditions not observed in the local Universe.  Recently, in order to study fainter galaxies with lower SFRs, researchers have focused their attention to galaxies which have been magnified by a strong gravitational lens, i.e. galaxy cluster potential well. This technique has allowed the study of galaxies and clump properties at spatial scales of a few hundreds of pc, otherwise prohibitive \citep[e.g.][]{2007MNRAS.376..479S, 2008Natur.455..775S, 2010MNRAS.404.1247J, 2012arXiv1209.5741L}. The most important outcome of this high-resolution analysis is that in spite of the extreme conditions under which clumps and star formation happens in these galaxies, the underlying physical process which governs star formation is the same as observed in the local Universe \citep[e.g.][]{2010Natur.464..733S}.

In this work, we will present an analysis of the clump population of a spiral galaxy at $z\sim 1.5$. Following the nomenclature by \citet{2011ApJ...732L..14Y}, we will refer to it as \object{Sp 1149}. The system is located behind the galaxy cluster \object{MACS J1149.5+2223} (hereafter MACS J1149), and thanks to the gravitational magnification, allows us to resolve spatial scales down to $\sim 100$ pc. According to \citet{2009ApJ...707L.163S}, Sp\,1149 has a luminosity comparable to a local $L^*$ galaxy \citep{2002MNRAS.336..907N}, and is about 1 mag fainter than previously studied unlensed spirals at similar redshifts. We will try to constrain how star formation is proceeding in the galaxy, by means of the clump analysis. The current SFR of Sp\,1149 is not extreme if compared to local spirals \citep[in the range 1--6 $\msun$/yr; ][]{2009ApJ...707L.163S, 2012arXiv1209.5741L}. \citet{2011ApJ...732L..14Y} reported clear signs of a rotationally supported disk and an increasing radial metallicity gradient toward the centre. Therefore, Sp\,1149 is a potential target for investigating the transition phase that extremely star-forming galaxies have experienced when evolving into modern late type systems. 

\begin{figure*}
\includegraphics[scale=0.4]{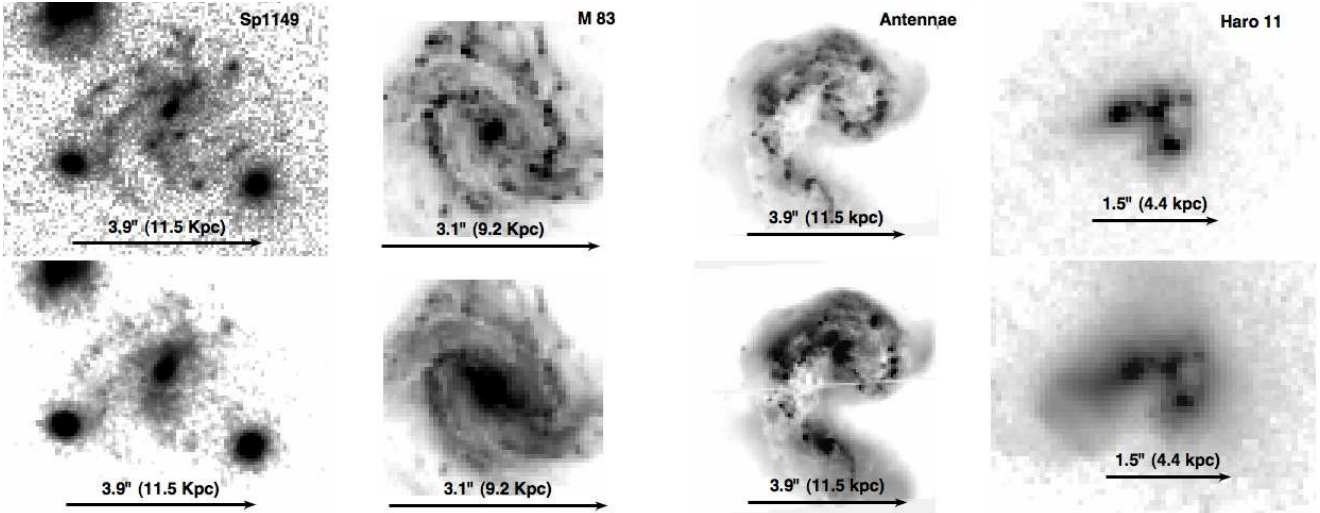}
\caption{Composite of $U$ (top) and $V$ (bottom) rest--frame images at $z=1.49$ of Sp\,1149, M\,83, Antennae, and Haro\,11. The Sp\,1149 frames correspond to the observed WFC3/F814W and WFC3/F125W imaging at pixel scale of 0.065''/px, respectively. The position of the clumps has been outlined in Figure~\ref{pos}. Please, notice that the three bright systems (one North-Est and two South-Est and West) are on the line of sight of this lensed image, but are not part of Sp\,1149. The M\,83 (VLT/ FORS-2 $U$ and $V$ bands, courtesy of S. Larsen), Antennae (WFC3/F336W and WFC3/F555W), Haro\,11 (ACS/F330W and ACS/F550M) data have been rescaled (in pixel scales and count rates) to match the magnified resolution and luminosity of Sp\,1149. In Section~\ref{sim-highz}, we will discuss the morphology of this simulated sample with respect to the Sp\,1149 clumpy morphology. The images are shown in logarithmic scales and rotated north-up. The angular--size distances at $z=1.49$ takes into account the square root of  the magnification factor (see main text).  \label{image}}
\end{figure*}

The paper is organised as follow. In Section 2, we introduce the dataset and details of the photometric analysis. These data will be used to derive clump properties, e.g. age, mass, extinctions, effective radii, in Section 3. In Section 4, we introduce some new data, together with literature ones, which will be used as a comparative sample, both at high $z$ and in the local Universe. Section 5 presents the analysis of the clump properties, in the context of local cluster complexes but also high $z$ clumps. Finally, in Section 6 we discuss the formation and fate of Sp\,1149 clumps, evidence for the bulge formation and the potential rate of metal-rich globular cluster progenitors. A discussion and summary will follow in the final section.

Throughout the paper we assume the following cosmological parameters: H$_{0}=73$ km s$^{-1}$ Mpc$^{-1}$, $\Omega_{M}=0.27$, and $\Omega_{\Lambda}=0.73$.

\section{Dataset \& photometric catalogue}

Our target is a lensed spiral galaxy with spectroscopically confirmed redshift $z=1.49$ \citep{2009ApJ...707L.163S}.  Four multiple images of \object{Sp 1149} have been identified in the galaxy cluster frame of MACS\,J1149. In this work we will focus on the tangential image A1.3 from \citet{2009ApJ...707L.163S} and showed here in Figure~\ref{image}. The particular geometry of the gravitational potential causes this image of \object{Sp 1149} to be almost equally lensed in both linear directions, thus avoiding the distortion effects that plague many other lensed galaxies. Using the lens model derived by \citet{2009ApJ...707L.163S}, we measured a linear magnification in the centre of the galaxy is 2.7$\times$3.1 (almost symmetric) at a position angle of 37 degrees  \citep[][Livermore et al in prep.]{2012arXiv1209.5741L}. Therefore, we assume a uniform magnification across the galaxy of $\mu = 8$ because the system is distant enough from the $z=1.5$ critical line to not be affected by strong magnification gradients. The uncertainties on the magnification as function of the position in the galaxy are smaller than the uncertainties associated to the computed galaxy cluster magnification map, i.e. $\sim 18$ \%. Because the fluxes of each clump are scaled by the same magnification factor, this uncertainty will not change the shape of the clump SED but only the derived mass.

The galaxy cluster MACS\,J1149 has been included among the 25 targets of the Cluster Lensing And Supernova survey with Hubble ({\it CLASH}), an ongoing {\it Hubble Space telescope} (HST) Multi-Cycle Treasury program (12065, PI: Marc Postman).  MACS\,J1149 has been imaged in 17 HST filters from the $UV$ to the $NIR$\footnote{released data available at http://archive.stsci.edu/prepds/clash/}. In Table~\ref{tbl-1A} and \ref{tbl-1B}, we summarise the filters used in this analysis (11 out of 17), the aperture correction applied, and the photometric catalogue of the extracted cluster complexes in the ABmag system. 

The CLASH--team has released two reduced datasets of MACS\,J1149 with different final pixel scale resolutions, 0.03''/px and 0.065"/px. The images have all been rescaled to the same pixel scale, aligned, and registered \citep[see][for more information]{2012ApJS..199...25P}. To carry out our analysis we use the set with 0.065"/px scale resolution. The latter is a good compromise between the finer resolution of the ACS/WFC and WFC3/UVIS cameras (0.05 and 0.04''/px, respectively) and the worse scale sampling of the WFC3/IR (0.13''/px). 

To perform the photometric analysis we selected only the filters where the signal--to--noise (S/N) was good enough to detect the clumps (e.g. in the F225W, F275W, and F336W filters the galaxy is barely detected, so they are excluded from the analysis) and the filter width was such that it carried new information (e.g. the F625W has been excluded because it overlaps with the F555W, F606W, and F775W transmission).  Of each large frame, we cut out a 13$\times$13 arcsec$^2$  image centred on the spiral. The ACS frames have then been combined with the {\tt IRAF} task {\tt imcombine}, using the function {\tt SUM} to improve the S/N of the sources. This combined frame has only been used to identify  the clumps by eye. To do aperture photometry, we used a radius of 3 px ($\sim0.2''$) and a sky annulus 1 px wide located at a distance of 4 px. The aperture radius has been chosen large enough to contain the stellar FWHM of the IR camera. The combined frame used for clump detection and the aperture size applied to do photometry is shown in Figure~\ref{pos}. 

As we will discuss in Section~\ref{ssc}, some of the clumps are extended, others unresolved or barely resolved. To compensate for the missing flux caused by the use of a finite aperture radius we used unsaturated stars detected in each initial MACS\,J1149 frame to estimate the aperture correction. In the case of extended sources this correction factor is a lower limit to the fraction of missed flux. In other words, we have only corrected for the missing flux assuming a stellar PSF. The photometry has also been corrected for Galactic extinction \citep{1998ApJ...500..525S}. 

Tables~\ref{tbl-1A} and \ref{tbl-1B} contain the final photometric catalogue of the 31 clumps. These fluxes have been scaled down by the factor $\mu=8$ to take into account the magnification of the lens. Photometry with a magnitude error larger than 1 mag has been discarded and considered as non--detection (empty cells in the tables). 

\begin{deluxetable*}{lccccccc}
\tabletypesize{\footnotesize}
\tablecaption{Part I. Photometric catalogue.\label{tbl-1A}}
\tablewidth{0pt}
\tablehead{
\colhead{ID}& \colhead{F390W} & \colhead{F475W} &
\colhead{F555W} & \colhead{F606W} & \colhead{F775W\tablenotemark{a}} &
\colhead{F814W\tablenotemark{b}} & \colhead{F850LP\tablenotemark{c}}
}
\startdata
Camera&WFC3&ACS&ACS&ACS&ACS&ACS&ACS\\
&UVIS&WFC&WFC&WFC&WFC&WFC&WFC\\
Exp--time (s)&4781&4136&9000&4128&4094&13548&8280\\
AC\tablenotemark{d}&-0.15&-0.16&-0.18&-0.16&-0.22&-0.33&-0.25\\
&&&&&&&\\
0&28.48$\pm$0.13&29.34$\pm$0.18&29.01$\pm$0.09&28.63$\pm$0.14&29.81$\pm$0.78&28.62$\pm$0.09&29.04$\pm$0.30\\
1&28.58$\pm$0.12&28.99$\pm$0.19&29.02$\pm$0.13&29.45$\pm$0.21&29.15$\pm$0.30&28.93$\pm$0.11&28.58$\pm$0.19\\
2&29.21$\pm$0.21&29.58$\pm$0.38&29.52$\pm$0.21&28.89$\pm$0.10&29.48$\pm$0.42&29.00$\pm$0.19& \nodata\\
3&29.16$\pm$0.15& \nodata&29.50$\pm$0.24&29.31$\pm$0.22& \nodata&28.80$\pm$0.16& \nodata\\
4&28.78$\pm$0.14&28.91$\pm$0.18&28.98$\pm$0.15&28.97$\pm$0.11&29.21$\pm$0.26&29.03$\pm$0.12&29.56$\pm$0.69\\
5&29.85$\pm$0.31&29.29$\pm$0.20&30.40$\pm$0.59&29.63$\pm$0.28&29.52$\pm$0.43&29.49$\pm$0.28&28.63$\pm$0.26\\
6&29.40$\pm$0.16&29.79$\pm$0.39&29.61$\pm$0.26& \nodata&29.31$\pm$0.34&29.47$\pm$0.17&29.85$\pm$0.59\\
7&29.50$\pm$0.26&29.86$\pm$0.31&29.30$\pm$0.17& \nodata&30.40$\pm$0.90&29.32$\pm$0.14&29.22$\pm$0.56\\
8&29.57$\pm$0.26&29.45$\pm$0.23&29.25$\pm$0.12&29.33$\pm$0.11&30.33$\pm$0.82&29.41$\pm$0.20& \nodata\\
9&29.50$\pm$0.33& \nodata&29.13$\pm$0.10&29.51$\pm$0.23& \nodata&30.91$\pm$0.72&29.47$\pm$0.61\\
10&29.46$\pm$0.40&29.58$\pm$0.43& \nodata& \nodata&29.93$\pm$0.77&29.74$\pm$0.20& \nodata\\
11&29.36$\pm$0.26&30.27$\pm$0.76&29.64$\pm$0.27& \nodata&29.83$\pm$0.56&29.70$\pm$0.27&29.33$\pm$0.59\\
12& \nodata&29.34$\pm$0.24&28.98$\pm$0.12&30.11$\pm$0.33&29.61$\pm$0.52&29.19$\pm$0.15&28.58$\pm$0.23\\
13&29.11$\pm$0.21&29.05$\pm$0.25&29.99$\pm$0.62&30.16$\pm$0.48&28.78$\pm$0.26&28.38$\pm$0.14&28.92$\pm$0.36\\
14&29.15$\pm$0.18&29.22$\pm$0.28&28.75$\pm$0.13&29.27$\pm$0.22&29.07$\pm$0.33&28.62$\pm$0.11&29.63$\pm$0.65\\
15&30.06$\pm$0.54& \nodata&30.10$\pm$0.60&29.79$\pm$0.38&29.98$\pm$0.60&29.50$\pm$0.25& \nodata\\
16&29.70$\pm$0.30&29.53$\pm$0.27&29.97$\pm$0.43&30.37$\pm$0.80&29.56$\pm$0.38&29.48$\pm$0.22& \nodata\\
17&30.30$\pm$0.59&29.61$\pm$0.38& \nodata& \nodata&29.56$\pm$0.38&29.80$\pm$0.35& \nodata\\
18&30.08$\pm$0.69&29.64$\pm$0.36&29.48$\pm$0.31&29.41$\pm$0.34&29.16$\pm$0.43&28.70$\pm$0.22&28.41$\pm$0.23\\
19&29.51$\pm$0.25& \nodata&29.96$\pm$0.15&30.00$\pm$0.38&30.05$\pm$0.82&29.17$\pm$0.13&29.62$\pm$0.82\\
20& \nodata&29.62$\pm$0.29& \nodata&30.07$\pm$0.26& \nodata& \nodata& \nodata\\
21& \nodata& \nodata& \nodata& \nodata&29.57$\pm$0.41& \nodata&30.06$\pm$0.95\\
22&31.30$\pm$0.97&30.61$\pm$0.91& \nodata&31.10$\pm$0.96& \nodata&31.26$\pm$0.93& \nodata\\
23&30.75$\pm$0.79& \nodata&31.01$\pm$0.87&30.89$\pm$0.84&28.52$\pm$0.23&29.71$\pm$0.24&29.24$\pm$0.55\\
24&30.75$\pm$0.66&29.67$\pm$0.41& \nodata&29.68$\pm$0.23& \nodata& \nodata& \nodata\\
25& \nodata&29.66$\pm$0.45&29.25$\pm$0.16&29.50$\pm$0.28& \nodata&29.66$\pm$0.26&29.65$\pm$0.80\\
26&30.09$\pm$0.32&29.72$\pm$0.44&31.05$\pm$0.62&30.57$\pm$0.59&29.19$\pm$0.24&30.35$\pm$0.33& \nodata\\
27&30.24$\pm$0.55&29.42$\pm$0.27&29.20$\pm$0.13&29.81$\pm$0.31& \nodata&29.61$\pm$0.16&29.36$\pm$0.52\\
28& \nodata& \nodata&29.63$\pm$0.21&30.08$\pm$0.42&29.05$\pm$0.22&29.74$\pm$0.26& \nodata\\
29 (bulge)&27.63$\pm$0.06&27.44$\pm$0.04&27.41$\pm$0.03&27.33$\pm$0.04&26.96$\pm$0.05&26.96$\pm$0.02&26.95$\pm$0.07\\
30&29.97$\pm$0.43& \nodata&29.97$\pm$0.23&30.23$\pm$0.43& \nodata&30.50$\pm$0.46&30.18$\pm$0.95\\
\enddata


\tablecomments{The filter throughput is in ABmag system. The listed magnitudes have been corrected for a constant factor (2.26 mag) to remove the magnification effect.}

\tablenotetext{a}{After $k$--correction, this filter is used as rest-frame $U$ band in the comparison with low-$z$ cluster complexes.} 
\tablenotetext{b}{main spectral features transmitted, [O{\sc ii}].} 
\tablenotetext{c}{main spectral features transmitted, Balmer--break, [O{\sc ii}].}
\tablenotetext{d}{Applied aperture correction, see main text for details.}
\end{deluxetable*}

\begin{deluxetable}{lcccc}
\tabletypesize{\footnotesize}
\tablecaption{PartII. Photometric catalogue.\label{tbl-1B}}
\tablewidth{0pt}
\tablehead{
\colhead{ID} & \colhead{F105W\tablenotemark{e}} & \colhead{F110W\tablenotemark{f}} & \colhead{F125W\tablenotemark{g}}&\colhead{F160W\tablenotemark{h}}
}
\startdata
Camera&WFC3&WFC3&WFC3&WFC3\\
& IR & IR& IR & IR\\
Exp--time [s]&2814.7&2414.7&2514.7&5029.3\\
AC&-0.38&-0.39&-0.45&-0.60\\
&&&&\\
0&28.73$\pm$0.10&28.43$\pm$0.07&28.41$\pm$0.06&28.70$\pm$0.12\\
1&29.34$\pm$0.15&28.80$\pm$0.08&29.17$\pm$0.16&29.27$\pm$0.07\\
2& \nodata&29.29$\pm$0.45&29.35$\pm$0.57&28.17$\pm$0.16\\
3&28.88$\pm$0.27&28.53$\pm$0.08&28.72$\pm$0.28&28.62$\pm$0.29\\
4&28.85$\pm$0.11&29.03$\pm$0.13&29.26$\pm$0.15&29.18$\pm$0.23\\
5&29.38$\pm$0.19&29.61$\pm$0.31&29.23$\pm$0.21&29.21$\pm$0.30\\
6&30.06$\pm$0.25&30.03$\pm$0.41&30.57$\pm$0.50&29.36$\pm$0.19\\
7&29.48$\pm$0.30&29.22$\pm$0.26&29.13$\pm$0.28&29.77$\pm$0.49\\
8&29.73$\pm$0.34&29.54$\pm$0.17& \nodata&30.08$\pm$0.53\\
9&30.49$\pm$0.63&29.39$\pm$0.27&30.33$\pm$0.77&29.55$\pm$0.53\\
10&30.15$\pm$0.39&29.51$\pm$0.24&29.44$\pm$0.15& \nodata\\
11&29.06$\pm$0.35&29.25$\pm$0.25&28.63$\pm$0.17&28.47$\pm$0.17\\
12&28.87$\pm$0.24&28.72$\pm$0.21&28.74$\pm$0.31&27.94$\pm$0.17\\
13&27.75$\pm$0.07&27.72$\pm$0.07&27.48$\pm$0.08&27.22$\pm$0.10\\ 
14&28.42$\pm$0.11&28.19$\pm$0.04&28.14$\pm$0.10&27.90$\pm$0.09\\
15&29.21$\pm$0.22&28.54$\pm$0.09&28.60$\pm$0.17&27.99$\pm$0.14\\
16&28.86$\pm$0.18&29.26$\pm$0.32&28.91$\pm$0.35& \nodata\\
17&29.81$\pm$0.36& \nodata& \nodata& \nodata\\
18&27.67$\pm$0.08&27.90$\pm$0.10&27.67$\pm$0.11&27.33$\pm$0.11\\
19&29.93$\pm$0.22&29.19$\pm$0.13&29.82$\pm$0.23&29.35$\pm$0.18\\
20&29.79$\pm$0.32&30.67$\pm$0.64&29.61$\pm$0.26&30.34$\pm$0.74\\
21&30.42$\pm$0.73& \nodata& \nodata& \nodata\\
22&29.77$\pm$0.33& \nodata& \nodata& \nodata\\
23&29.20$\pm$0.17&29.46$\pm$0.25&28.99$\pm$0.17& \nodata\\
24& \nodata&30.30$\pm$0.46& \nodata& \nodata\\
25&29.26$\pm$0.19&29.45$\pm$0.30&29.08$\pm$0.26&29.18$\pm$0.33\\
26&29.40$\pm$0.19&30.03$\pm$0.42&29.93$\pm$0.35&29.19$\pm$0.14\\
27&29.75$\pm$0.18&29.20$\pm$0.13&29.91$\pm$0.31&30.51$\pm$0.66\\
28&30.20$\pm$0.74&28.89$\pm$0.16& \nodata& \nodata\\
29 (bulge)&26.65$\pm$0.02&26.65$\pm$0.03&26.55$\pm$0.03&26.25$\pm$0.04\\
30&30.41$\pm$0.41&29.85$\pm$0.14&29.45$\pm$0.19&30.40$\pm$0.41\\
\enddata


\tablecomments{Continue from Table~\ref{tbl-1A}.}

\tablenotetext{e}{main spectral features transmitted, Balmer--break, [O{\sc ii}].} 
\tablenotetext{f}{main spectral features transmitted, Balmer--break, [O{\sc ii}], [H$\beta$], [O{\sc iii}].} 
\tablenotetext{g}{main spectral features transmitted, H$\beta$, [O{\sc iii}]. After $k$--correction, this filter is used as rest-frame $V$ band in the comparison with low-$z$ cluster complexes.} 
\tablenotetext{h}{main spectral features transmitted, H$\alpha$.}
\end{deluxetable}

\section{Sp\,1149 clump analysis}

\subsection{Determining age and mass of the cluster complexes}

The multi-band photometry has been used to perform a SED analysis of each clump. 

The evolutionary tracks have been produced with the publicly available Yggdrasil models\footnote{http://ttt.astro.su.se/$\sim$ez} \citep{2011ApJ...740...13Z}. We assume two different metallicities Z$=$0.004 and 0.008, Starburst99 single stellar population (SSP) models based on Padova-AGB tracks  \citep{1999ApJS..123....3L, 2005ApJ...621..695V} and a universal Kroupa (2001) IMF in the interval 0.1--100 $\msun$. To regulate the relative contribution from nebular emission to the final SED we assumed a covering factor, $f_{cov}=1$. This assumption also determines the Lyman continuum escape fraction, $f_{esc}$ through the relation $f_{esc} =1- f_{cov}$, i.e. none in our case. This choice of parameters seems reasonable if one considers that the radius inside which we do photometry corresponds to a projected physical size of about 600 pc, i.e. large enough to have a high probability for the ionising photons to be absorbed. Finally, we have simulated three star formation histories (SFHs), instantaneous burst (e.g. SSP), constant star formation rate (SFR) for 10 Myr, and for 100 Myr (i.e. it produces a burst of star formation with constant SFR lasting a given interval of Myr, after which the SFR drops to zero).

The $\chi^2$-algorithm used is described in \citet{2010MNRAS.407..870A}. In short, comparing models and observed fluxes, the algorithm gives a best solution for a combination of age, internal extinction, and mass. We allowed data-points with photometric errors as large as 1 mag to be included in the fit (notice that the weight on the fit of these data is, therefore, very small). We have implemented the algorithm with the analysis of 68 \% confidence limits on the determination of the clump parameters (associated errors in Table~\ref{tbl-3}). According to the prescription given by \citet{1976ApJ...208..177L}, the uncertainties associated to the best outcome parameters can be determined if the numbers of fitted data points is larger than the number of parameters (e.g. three in our case).  In our catalogue, only source \#21 did not pass this criterion (detected in only three passbands), therefore we could not perform the error analysis.

\begin{deluxetable}{lcccccc}
\tabletypesize{\footnotesize}
\tablecaption{Physical properties of  Sp\,1149 cluster complexes.\label{tbl-3}}
\tablewidth{0pt}
\tablehead{
\colhead{ID} & \colhead{M$_U$} & \colhead{M$_V$} &\colhead{R$_{eff}$ [pc]}& \colhead{Age [10 Myr]}&\colhead{Mass [$10^7 \msun$]}& \colhead{E(B-V) [mag]}
}
\startdata
0&-15.52&-15.74&$<$120&3.00$^{+0.00}_{-2.00}$ &1.07$^{+0.21}_{-0.88}$ &0.04$^{+0.11}_{-0.00}$ \\
1&-16.20&-15.02&$<$120&1.00$^{+0.00}_{-0.50}$ &0.08$^{+-0.00}_{-0.03}$ &0.01$^{+0.01}_{-0.00}$ \\
2&-15.91&-14.86& \nodata&0.10$^{+2.00}_{-0.00}$ &0.10$^{+0.99}_{-0.03}$ &0.15$^{+0.24}_{-0.00}$ \\
3& \nodata&-15.42& $<$120&6.00$^{+3.00}_{-5.90}$ &1.38$^{+0.46}_{-1.28}$ &0.01$^{+0.24}_{-0.00}$ \\
4&-16.15&-14.94&$<$120&1.00$^{+2.00}_{-0.50}$ &0.10$^{+0.53}_{-0.03}$ &0.02$^{+0.06}_{-0.00}$ \\
5&-15.84&-14.96&$<$120&1.00$^{+2.00}_{-0.90}$ &0.11$^{+0.54}_{-0.07}$ &0.17$^{+0.18}_{-0.00}$ \\
6&-16.06&-13.63&$<$120&0.50$^{+0.50}_{-0.50}$ &0.04$^{+0.02}_{-0.01}$ &0.01$^{+0.06}_{-0.00}$ \\
7&-14.93&-15.02&$<$120&3.00$^{+0.00}_{-3.00}$ &0.54$^{+0.48}_{-0.50}$ &0.00$^{+0.21}_{-0.00}$ \\
8&-14.97& \nodata&$<$120&2.00$^{+0.00}_{-1.40}$ &0.22$^{+0.09}_{-0.18}$ &0.03$^{+0.06}_{-0.00}$ \\
9& \nodata&-13.87& \nodata&1.00$^{+0.00}_{-0.50}$ &0.06$^{+0.01}_{-0.02}$ &0.00$^{+0.06}_{-0.00}$ \\
10&-15.47&-14.79& \nodata&0.20$^{+7.80}_{-0.20}$ &0.07$^{+0.76}_{-0.03}$ &0.00$^{+0.36}_{-0.00}$ \\
11&-15.57&-15.60&122&0.20$^{+2.80}_{-0.20}$ &0.16$^{+2.37}_{-0.10}$ &0.17$^{+0.38}_{-0.02}$ \\
12&-15.75&-15.45&122&1.00$^{+19.05}_{-1.00}$ &0.48$^{+3.55}_{-0.43}$ &0.17$^{+0.43}_{-0.00}$ \\
13&-16.69&-16.70&2439&20.00$^{+0.00}_{-17.05}$ &14.00$^{+4.30}_{-5.01}$ &0.14$^{+0.43}_{-0.06}$ \\
14&-16.26&-16.01&244&3.00$^{+0.00}_{-2.90}$ &2.73$^{+0.53}_{-2.51}$ &0.18$^{+0.27}_{-0.11}$ \\
15&-15.42&-15.63&$<$120&0.10$^{+1.40}_{-0.00}$ &0.32$^{+1.69}_{-0.15}$ &0.34$^{+0.50}_{-0.19}$ \\
16&-15.83&-15.23&244&8.00$^{+2.02}_{-8.00}$ &1.23$^{+1.19}_{-1.17}$ &0.00$^{+0.39}_{-0.00}$ \\
17&-15.75& \nodata&122&1.50$^{+8.52}_{-1.50}$ &0.18$^{+1.22}_{-0.16}$ &0.08$^{+0.47}_{-0.00}$ \\
18&-16.31&-16.51&976&20.05$^{+0.00}_{-17.05}$ &11.80$^{+3.50}_{-3.62}$ &0.12$^{+0.39}_{-0.06}$ \\
19&-15.35&-14.42&487&0.10$^{+0.90}_{-0.10}$ &0.04$^{+0.07}_{-0.00}$ &0.04$^{+0.14}_{-0.00}$ \\
20& \nodata&-14.53& \nodata&1.50$^{+2.50}_{-1.40}$ &0.07$^{+0.24}_{-0.05}$ &0.00$^{+0.14}_{-0.00}$ \\
21&-15.80& \nodata& \nodata&0.50&0.03&0.00\\
22& \nodata& \nodata& \nodata&30.23$^{+872.57}_{-30.23}$ &1.85$^{+7.8e+6}_{-1.83}$ &0.00$^{+3.00}_{-0.00}$ \\
23&-16.81&-15.16&244&3.00$^{+17.05}_{-3.00}$ &2.01$^{+2.40}_{-1.91}$ &0.45$^{+0.62}_{-0.00}$ \\
24& \nodata& \nodata& \nodata&1.40$^{+3.60}_{-1.40}$ &0.10$^{+0.35}_{-0.08}$ &0.11$^{+0.23}_{-0.00}$ \\
25& \nodata&-15.06&487&3.00$^{+0.00}_{-3.00}$ &0.51$^{+0.35}_{-0.46}$ &0.00$^{+0.18}_{-0.00}$ \\
26&-16.20&-14.28&244&0.10$^{+10.02}_{-0.00}$ &0.05$^{+1.21}_{-0.02}$ &0.19$^{+0.38}_{-0.00}$ \\
27& \nodata&-14.28&122&1.00$^{+2.00}_{-0.90}$ &0.06$^{+0.38}_{-0.03}$ &0.03$^{+0.10}_{-0.00}$ \\
28&-16.32& \nodata&1707&6.00$^{+14.05}_{-6.00}$ &0.91$^{+2.61}_{-0.87}$ &0.13$^{+0.55}_{-0.00}$ \\
29 (bulge)&-18.37&-17.60&$<$120&3.00$^{+0.00}_{-2.90}$ &11.59$^{+2.84}_{-10.61}$ &0.16$^{+0.28}_{-0.11}$ \\
30& \nodata&-14.70&244&3.00$^{+0.00}_{-2.90}$ &0.25$^{+0.09}_{-0.23}$ &0.00$^{+0.11}_{-0.00}$ \\
\enddata
\tablecomments{The rest-frame absolute magnitudes in $U$ and $V$ bands have been obtained applying a $k$-correction to the F775W and F125W filters and converted from ABmag to Vega system to allow the comparison with the local reference sample. The ages, masses, and extinction have been derived using models with 10 Myr continuos star formation. See main text for details.}
\end{deluxetable}

\begin{figure}
\includegraphics[scale=0.45]{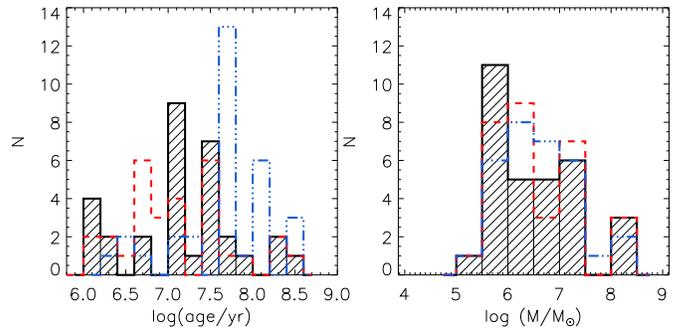}
\caption{Recovered clump age (left) and mass (right) distributions. The outputs have been obtained for three sets of {\it Yggdrasil models} and metallicity Z$=0.008$; instantaneous burst (red dashed line histogram), continuous SFR for 10 Myr (black solid line histogram, these outcomes will be used in the analysis), continuous SFR for 100 Myr (blue dotted-dashed line histogram).  \label{cl_prop}}
\end{figure}

In Figure~\ref{cl_prop}, we show the recovered age and mass distributions of the clumps. We only use the outcomes of the fit obtained with the three SFHs and metallicity Z$=$0.008 models. We notice, however, that the fit with models at lower metallicity (Z$=$0.004) produces similar results. We preferred the former metallicity assumption because it is closer to the mean value found by \citet{2011ApJ...732L..14Y}.  The SFH of the clumps is unknown. If we look at cluster complexes in the local Universe (see last column in Table~\ref{tbl-4}) their age range variates from a few Myr (in agreement with an assumption of instantaneous burst) to a few tens of Myrs (compatible with prolonged star formation episodes). Therefore, we decided to compare the results obtained from the 3 different SFHs, instantaneous, continuous for 10 Myr (similar to the range observed for more massive and complex star-forming knots), and for 100 Myr. Most likely the real SFH lies between these two last values. In Figure~\ref{cl_prop}, the age distribution recovered with an instantaneous burst model (red dashed line histogram) produces a peak at young ages ($\sim 6-7$ Myr) which disappears if a continuous SFR (a more realistic assumption) for a given time model is used, instead. The ages we fit are defined from the beginning of the each star formation episode of each model. Allowing for a range of star formation histories introduces a certain age degeneracy because blue objects can be fitted to older ages for more prolonged star formation scenarios, but since the overall SED is continuously evolving even if the SFR stays constant \citep[e.g.][]{2005A&A...435...29Z}, many of our clumps turn out to be best-fitted by ages shorter than the assumed duration of the constant-SFR phase. Models with prolonged SFR have, as expected, the tendency to produce slightly older ages for the clumps. On the other hand, the mass distributions do not change significantly  and range between a few times $10^5$ up to $10^8 \msun$ (more than 70 \% have masses below $3\times10^7 \msun$), in spite of the different models used. The recovered mass range is in good agreement with the results obtained by \citet{2009ApJ...692...12E}.  We decided to use the results obtained with the fit to the model of 10 Myr continuous star formation and metallicity Z$=$0.008 in the rest of the analysis. However, we stress that the conclusions would not change significantly if the results obtained with a 100 Myr SFH history would have been used, instead.

\subsection{Clump size} \label{ssc}
To measure the clump size we used the frame with the deepest exposure, i.e. the F814W. Any attempt to directly measure the FWHM of the Sp\,1149 clumps has failed because the S/N is not high enough except for a few sources, e.g. \#1, \#4, and the bulge (\#29). For these sources we measured FWHM (2.67, 2.53, 2.61 px, respectively) comparable to the mean FWHM of the stellar PSF ($<\sigma>=2.64\pm0.33$ px). 

\begin{figure}
\includegraphics[scale=0.45]{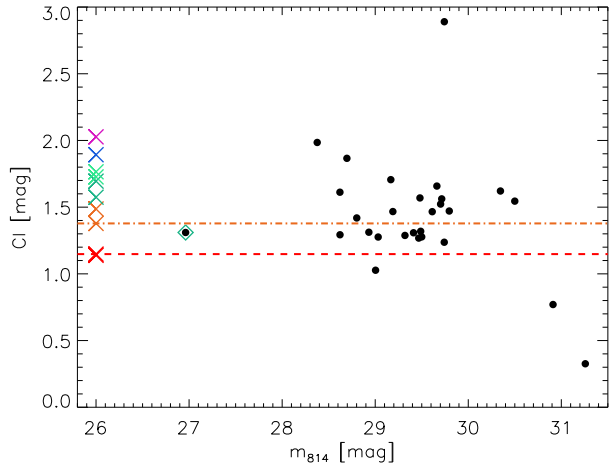}
\caption{Observed concentration indexes (CI) as function of their final (de-lensed) F814W magnitude (black dots). On the left side the crosses indicates the reference CI values used to derive the R$_{eff}$ of the clumps. The red cross correspond to the stellar CI.  The other crosses, from orange to purple shows the recovered CI for increasing values of R$_{eff}$ of simulated extended sources. Sources below the red dashed line are unresolved. The objects which fall between the dashed and dot-dashed lines have upper-limits to their sizes. Sources above the dot-dashed line are resolved. The green diamond shows which source is the bulge of the galaxy. See main text for details.  \label{CIdx}}
\end{figure}

To determine the sizes we applied an indirect deconvolution of the object PSF with the stellar PSF, which gives an approximate measure of the R$_{eff}$ \citep[the half-light radius, i.e. the radius that contains half of the total light of the system, e.g.][]{2010ARA&A..48..431P}.  

We looked at the concentration index (CI) of each clump, e.g. the concentration of the light as function of the PSF radius. We defined the CI as the difference between the magnitude of the clump at a photometric aperture of 1 px and the magnitude at a radius of 3 px. In Figure~\ref{CIdx}, we see the recovered values of CI as function of their final magnitude in the F814W filter (black dots). CI values close to 0.0 means that all the light of the source is concentrated in the central circle of 1 px radius.  The stellar PSF CI sets the limit between resolved and unresolved objects in the frame. We used a few stars in the F814W frame to measure this limit, obtaining a CI$=1.15$ mag (red cross in Figure~\ref{CIdx}). Clumps with CI below this limit are considered unresolved. 

To trace the behaviour of the CI as function of the size of the source, we estimated the CI of simulated extended objects with increasing R$_{eff}$. We constructed for the F814W filter a modelled PSF using the {\tt PyRAF} task {\tt PSF}. The stellar PSF was then convolved with an analytic surface profile \citep[][and power index of 3.0]{1987ApJ...323...54E} of increasing R$_{eff}$ using the task {\tt mkcmppsf} from the publicly available software {\tt BAOlab}\footnote{http://baolab.astroduo.org/} \citep{1999A&AS..139..393L}. Each extended profile has been used to include several mock sources in the F814W frame, for which we derived a median CI value. We used the following grid of R$_{eff}$ values, 0.06, 0.6, 1.3, 2.5, 3.8, 5.1, 6.4, 8.9, and 12.7 px (1 px $=$ 191 pc at the distance of Sp\,1149). The recovered CI values are shown as crosses in Figure~\ref{CIdx}, in a scale from orange to purple for increasing R$_{eff}$. The assigned size to each clump corresponds to the R$_{eff}$ of the simulated CI which is the closest to the observed one. Clumps with CI smaller than the stellar CI (red cross) or photometric errors larger than 1 mag have null measurements. The smallest recovered R$_{eff}$ we consider significant is R$_{eff}=0.6$ px$\sim122$ pc.  Sources with CI larger than the stellar one but smaller than half a pixel (R$_{eff} < 0.5$ px) have assigned an upper limit to their measurement corresponding to $\sim120$ pc. One can see that the clumps for which the FWHM has been directly measured on the image (\#1, 4 and 29) fall in this category (e.g. sources with an upper-limit to the real size). The recovered R$_{eff}$ are all listed in Table~\ref{tbl-3}. In a recent work, \citet{2012arXiv1209.5741L}, detected the bulge of Sp\,1149 in a H$\alpha$-dedicated survey with {\it HST}. They measured a size for our source \#29 of 174$\pm$34 pc, which is in good agreement with our determined upper limit of 120 pc, considering that the two measurements have been taken at different wavelengths (H$\alpha$ is generally more physically  extended than $U$-band rest-frame emission).

\section{Building a reference sample} \label{local}
In order to compare the properties of Sp\,1149 clumps to local and high $z$ complexes, we assembled a comprehensive reference sample which is used in Section~\ref{comp}. 

\begin{deluxetable*}{lccccc}
\tabletypesize{\footnotesize}
\tablecaption{Local reference sample properties\label{tbl-4}}
\tablewidth{0pt}
\tablehead{
\colhead{ID} & \colhead{M$_U$ [mag]}& \colhead{M$_V$ [mag]} & \colhead{R$_{eff}$ [pc]}&\colhead{Mass [$10^5 \msun$]}&\colhead{Age [Myr]\tablenotemark{a}}
}
\startdata
\cutinhead{Literature Data}
NGC\,6946-1\tablenotemark{b}&-15.81&-15.00&390.0&100.00&1--30\\
Antennae\tablenotemark{c}-1&-14.85&-13.80&140.0&22.50&20--30\\
Antennae-2&-14.69&-13.60&198.0&8.00&6--16\\
Antennae-3&-14.69&-13.60&200.0&8.00&7--14\\
Antennae-4&-15.08&-13.50&136.0&10.00&3.5--5.5\\
Antennae-5&-15.77&-14.00&175.0&11.00&3--4\\
Antennae-6&-14.85&-13.10&90.0&10.00&1--2\\
Antennae-1Nuc&-16.04&-14.60&123.0&9.00&4--6\\
Antennae-2Nuc&-16.07&-14.30&140.0&11.00&3.5\\
M\,51\tablenotemark{d}-C2&-13.40&-12.60&160.0&1.40&7\\
M\,51-D2&-13.20&-12.40&160.0&1.10&7\\
M\,51-E2&-12.00&-11.20&85.0&0.40&7\\
M\,51-F2&-11.90&-11.10&100.0&0.30&7\\
M\,51-G2&-14.10&-13.30&220.0&2.60&7\\
M\,51-B1&-12.80&-12.00&125.0&0.70&7\\
\cutinhead{This Work}
M\,83-0&-12.89&-11.56&50.0&0.44&1--3\\
M\,83-1&-14.12&-12.87&65.0&1.45&5--10\\
M\,83-2&-14.12&-12.86&70.0&1.43&6\\
M\,83-3&-13.55&-12.26&85.0&0.83&1--13\\
M\,83-4&-13.88&-12.58&110.0&1.12&1--14\\
M\,83-5&-13.86&-12.27&117.5&0.84&5--10\\
M\,83-6&-13.51&-12.23&117.5&0.80&5--10\\
M\,83-7&-13.20&-11.91&80.0&0.60&5\\
M\,83-8&-13.74&-12.36&120.0&0.91&1--10\\
M\,83-9&-13.06&-11.69&100.0&0.49&5\\
Antennae-Knot S\tablenotemark{e}&-17.25&-16.3&490 (20)&40&6-30\\
NGC\,7252\tablenotemark{f}-W3&-20.19(-16.2)&-19.10(-15.67)&17.5&960.0(800.0)&10(400)\\
NGC\,7252-W30&-18.39(-14.6)&-17.30(-14.07)&9.3&190.0(160.0)&10(400)\\
Haro\,11-Knot C&-19.35&-18.34&400.0&420&3--35\\
Haro\,11-Knot B&-18.42&-17.58&400.0&210&1--20\\
Haro\,11-Knot A&-18.79&-17.39&400.0&170&1--20\\
ESO\,338-knotA&-18.00&-16.67&301.0&27.41&3--11\\
\enddata


\tablecomments{The photometry is in Vega system. The distance modulus used for the new data are 28.25 mag for M\,83, 31.71 mag for the Antennae, 34.58 mag for Haro\,11, and 32.87 for ESO\,338.  }

\tablenotetext{a}{For the literature data, we report the mean age listed in the corresponding papers. We list, instead, the minimum and maximum age of the clusters located in the complexes of M\,83, in Knot S of the Antennae, in the knots of Haro\,11 and ESO\,338.} 
\tablenotetext{b}{\citet{2002ApJ...567..896L}.} 
\tablenotetext{c}{The Antennae data are taken from \citet{2006A&A...445..471B}, except the $U$-band luminosity (obtained from the color produced by our models at the corresponding age) and the radius of the complexes 1 and 2 Nuc (corresponding to the aperture radius used to do photometry).} 
\tablenotetext{d}{The M\,51 data are taken from \citet{2005A&A...443...79B}, except the $U$-band luminosity (obtained from the color produced by our models at the corresponding age).}
\tablenotetext{e}{For the cluster analysis of Knot S and the star formation history see \citet{2010AJ....140...75W}. In this work, we use the size of the whole complex (490 pc). We indicate in bracket the R$_eff=20$ pc as reported by Bastian et al (submitted). See main text for details. }
\tablenotetext{f}{The observed values at the current age of the clusters from \citet{2006A&A...448..881B} are shown in brackets. This information have been used to extrapolate the luminosity in $U$ and $V$ bands, applying a M/L ratio at 10 Myr given by our models.}
\end{deluxetable*}

\subsection{Cluster complexes in nearby galaxies}
For the local sample we extended the data available in the literature with new contributions. They are all summarised in Table~\ref{tbl-4}. 

The cluster complex properties of the Antennae and M\,51 are from \citet[][respectively]{2006A&A...445..471B, 2005A&A...443...79B}. For two of the Antennae complexes, the sizes were not available, therefore, we assigned the size of the aperture radius used to do photometry. In both studies, only the $V$ band luminosity was available. We derived the corresponding $U$ band magnitude from the color information at the given age (mean age from the interval given for each of the Antennae complexes, mean age of 7 Myr for all the complexes of M51, as indicated in the paper). The properties of the massive cluster complex in NGC\,6946 are available from  \citet{2002ApJ...567..896L}. The sample also includes some of the most massive known intermediate age super star clusters (SSCs), W3 and W30, located in the post-merger galaxy NGC\,7252 \citep{2006A&A...448..881B}. We used the age and M$_V$ information reported in the paper to derive the cluster properties (e.g. M$_U$, M$_V$, mass) at the age of 10 Myr, using SSP models (color and mass-to-light ratio, M/L$_V$).

The new contribution to this local sample comes from cluster complexes in the M\,83 spiral, an extended star-forming knot in the Antennae, Knot S  \citep{2010AJ....140...75W}, and in two starburst blue compact galaxies Haro\,11 and ESO\,338-IG04 (hereafter, ESO\,338 ). The M\,83 photometric data (WFC3/F336W and WFC3/F555W) have been presented in \citet{2011MNRAS.417L...6B}. The photometry of Knot S has been performed on the latest WFC3/UVIS F555W and F336W acquired images (GO 11577, PI: Whitmore). The reduction of the ACS/F330W and F550M frames, used for the analysis of Haro\,11 complexes, is described in \citet{2010MNRAS.407..870A}. The description of the data (PC/F336W and ACS/F550M frames) used for the analysis of ESO\,338 can be found in \citet[][]{2009AJ....138..923O} and \"Ostlin et al (in prep.).  We did aperture photometry using a radius of 115 px ($\sim$100 pc), 115 px ($\sim$490 pc), 40 px ($\sim$400 pc), and 35 px ($\sim$300 pc), for M\,83, Knot S, Haro\,11, and ESO\,338, respectively. The fluxes have been corrected for Galactic extinction. No aperture correction has been included. The photometry listed in Table~\ref{tbl-4} is in the Vega system. 

The masses have been estimated using the $V$-band luminosity and the M/L ratio given by {\it Yggdrasil} models (instantaneous burst)  corresponding to a stellar population of 5 Myr for cluster complexes in M\,83 and the knot of Eso\,338, 6 Myr for Knot S, 10 Myr for the knots in Haro\,11. We notice that these mass estimates have to be considered a lower limit to the real values. For consistency with the data available in the literature we use M/L ratio derived from SSP models at a single age. However, the SFH in these regions is more complex and the M/L ratio is very sensitive to the age of the stellar population. The deviations are important for more complex regions such as the star-forming knots in starburst galaxies. For example, dynamical mass estimates of the knots in Haro 11 gives about a  factor 10 higher masses \citep[][Ostlin et al. submitted]{2012A&A...541A.115G}. 

Because the regions we looked at are resolved, we defined the size of each complex as the radius which included all the bright sources in the clustered region. In cases where the regions are elongated we defined semi-minor/major axes and used in the analysis the mean of the two values. For the two SSCs W3 and W30 we report the values of the R$_{eff}$. However, Bastian et al. (submitted) report that their surface profiles are much more extended and observe the same behaviour in Knot S. This region is much younger than W3 and W30 but it could be considered a progenitor analog of these SSCs. The R$_{eff}$ measured for Knot S is about 20 times smaller than the radius of the complex (see Table~\ref{tbl-4}). This discrepancy  is likely caused by the fact that  one of the most massive SSCs observed in the Antennae \citep{2010AJ....140...75W} is located in the centre of Knot S, dominating is light.

Morphologically, we noticed that in spiral galaxies some of the complexes are very diffuse, while others are centrally concentrated and much brighter, but do not extend as far \citep[e.g. see images showed in][]{2005A&A...443...79B, 2010AJ....140...75W}. The cluster complex in NGC\,6946 is quite peculiar. Unlike complexes in other spirals it is more extended and massive, hosting a very massive cluster and several other intermediate mass ones. Complexes in blue compact galaxies are more massive, extended and host a large population of clusters, suggesting denser clustering. These regions are real starburst knots \citep[e.g.][]{2010MNRAS.407..870A} with $\Sigma_{SFR}$ an order of magnitude higher than the values presented in \citet{2006A&A...445..471B, 2005A&A...443...79B}.

\subsection{The high $z$ sample}

The high $z$ sample includes contributions from 3 different observational samples and one simulation work. The first sample includes the properties of the clumps in 6 SINS \citep[SINFONI Integral field spectroscopy survey of $z \sim 2$ galaxies,][]{2009ApJ...706.1364F} targets. NICMOS imaging of these systems has allowed the derivation of the size and mass of their clumps \citep{2011ApJ...739...45F}. We note that the resolution of the NICMOS data has limited the measurement to clumps with sizes larger than $\sim$1.4 kpc (using the FWHM of the clumps), i.e. an order of magnitude larger than our limit of 120 pc. The stellar masses have been estimated from the luminosity of the clumps in the F160W band (corresponding to the $V$ band rest-frame), assuming the M/L ratio at the age derived from the $i_{F814W}-F160W$ color. \citet{2012MNRAS.422.3339W} reported the sizes and masses of the clumps of three targets belonging to the WiggleZ Dark Energy Survey. The sizes were measured in the H$\alpha$ redshifted frame, fitting the light profile of the clumps. The masses are dynamical estimates using the velocity dispersion of the H$\alpha$ emission line and the assumption of virial equilibrium. The third sample includes sizes (from isophotes) and masses (dynamically) derived from clumps of gravitationally lensed targets  at $z\sim2-3$ \citep{2010MNRAS.404.1247J}. Finally, we consider also the size and mass of simulated clumps formed by violent gravitational instability in high-z disc galaxies \citep{2012MNRAS.420.3490C}.

\section{Understanding the physical properties of Sp\,1149} \label{comp}

\subsection{Luminosity properties}

In order to compare the luminosity of the Sp\,1149 clumps to the brightness of local cluster complexes, we have applied the $k$-correction to the photometry in the F775W and F125W filters. We used {\it Yggdrasil} models at $z=0$ and the constrained ages to derive the correction factor for each clump, obtaining corresponding absolute magnitude in $U$ (F336W) and $V$ (F555W) bands at $z=0$.

The total apparent magnitude of Sp\,1149 in the F814W frame is $I \simeq 23.4 \pm 0.3$ mag  \citep{2009ApJ...707L.163S}. A comparison with the total luminosity produced by the clumps shows that, in spite of being a prominent feature of Sp\,1149, they contribute only 15 \% of the total luminosity in the rest-frame $U$-band. In other words, the bulk of the UV light in the galaxy is produced by a diffuse stellar component. A similar result has already been found by \citet{2011ApJ...739...45F} in a sample of $z\sim2$ galaxies.  Based on its UV properties, Sp\,1149 seems to agree quite well with the trend observed at high $z$.

\begin{figure}
\includegraphics[scale=0.45]{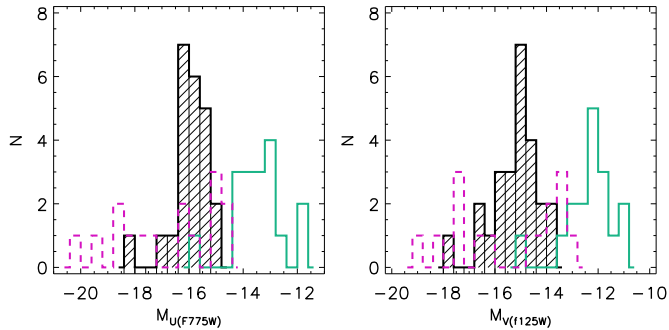}
\caption{The complex luminosity function in $U$ (left) and $V$ (right) band. The Sp\,1149 clump distribution is shown by the black shadowed histogram. The green continuous line shows the distribution of the cluster complexes in the spiral galaxies included in our sample (NGC\,6946, M\,51, M\,83). The purple dashed histogram shows the luminosity distribution of complexes in mergers and starburst galaxies (Antennae, NGC\,7252, Haro\,11, ESO\,338).  \label{lf_prop}}
\end{figure}

In Figure~\ref{lf_prop}, one can see the luminosity distribution of the clumps in Sp\,1149 compared to the luminosity distribution of the local reference sample. The luminosity of the clumps is similar to the range of brightness covered by the local sample. To better understand the similarities and differences between the two datasets, we divided the local sample into complexes belonging to spirals (green solid line distribution) and complexes formed in merger starburst systems (purple dashed line). The distribution of the $z\sim1.5$ clumps sits in between the 2 subsamples, about 2-3 magnitudes brighter than the mean magnitude of the complexes in spirals. One can see that the high-$z$ distribution partially overlaps with the  brightest systems in the Antennae (secondary peak in the purple distribution at lower brightness bins). Definitively, the Sp\,1149 clumps are not as luminous as the knots in blue compact galaxies or the most massive single clusters of NGC\,7252 (bright-end purple distribution). Only the bulge-like component of Sp\,1149 (the brightest source in the galaxy) has luminosity comparable to the star-forming knots. We notice that the median value of the Sp\,1149 clump luminosity is M$_{U}=-15.84$ mag and M$_V=-15.02$ mag. These median values overlap quite well with the observed magnitudes of the NGC\,6946 massive cluster complex (listed in Table~\ref{tbl-4}), one of the most massive studied complexes in local spirals. 

\begin{figure}
\includegraphics[scale=0.45]{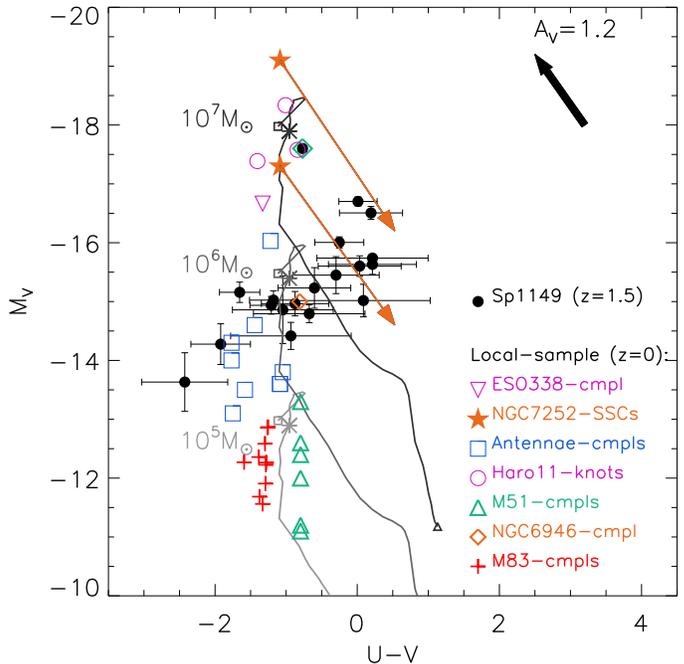}
\caption{A color--magnitude diagram of the clumps (black dots) and the local reference sample (symbols are labelled in the inset). The black arrow shows in which direction of the diagram the data--points would move after an extinction correction is applied. Photometric error bars are included. The errors of the local sample are very small (order of 0.005 mag) so are not shown. The green diamond enclosing the brightest black dot shows the position of the bulge-like source. The orange stars indicate the position of the two very massive star clusters of NGC\,7252 at the age of 10 Myr. The orange arrows show the current position, corresponding to the age of 400 Myr.  The over-plotted tracks are from {\it Yggdrasil} models at $z=0$, assuming Z$=0.008$, covering factor of 1, Kroupa IMF, and 10 Myr of constant SFR. Different stellar masses are assumed (labelled in the plot). A square and a triangle symbol indicates the starting and ending of the tracks, while the asterisk shows the position in the track corresponding to 10 Myr. \label{cmd}}
\end{figure}

\begin{figure}
\includegraphics[scale=0.4]{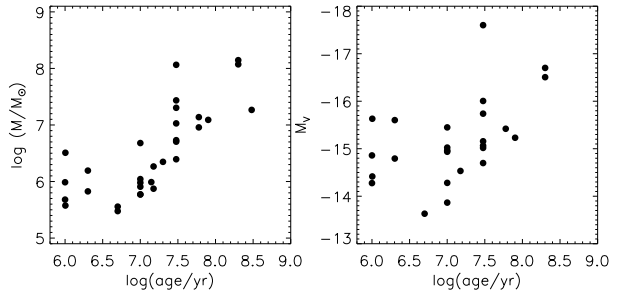}
\caption{Left: recovered ages versus masses of the Sp\,1149 clumps. Right: recovered ages versus absolute luminosity in $V$-band. The latter plot contains less objects because of a few non-detections in the F125W filter (the corresponding $V$-band rest-frame filter). \label{age-mass}}
\end{figure}

The trend observed in the luminosity function is better understood in a color--magnitude diagram (Figure~\ref{cmd}), in which the information coming from the two filters can be explored and correlated.   
In general, we observe that the high $z$ clumps are distributed in a wide range of $U-V$ color. Clumps with $U-V < -0.5$ mag overlap quite well with the range of $U-V$ color of the local cluster complexes, suggesting that they have similar young ages. These young clumps seem to be the ones with lower stellar mass and are comparable to the complexes in the Antennae or in NGC\,6946. On the other hand, clumps with  redder colors, e.g. $U-V>-0.5$ mag seem to be older and more massive than their younger counterparts. However, because of the extinction, some of these redder clumps could still be young. Using the derived ages we see that about 60 \% of the clumps have an age between 1-30 Myr (corresponding to $U-V=-0.5$). The remaining fraction is definitively older. The coexistence of two populations of clumps (young, low mass and less bright, versus old, more massive and bright) in Sp\,1149 is clearly outlined by the distribution of the clumps in the age--mass  and age--M$_V$ diagrams shown in Figure~\ref{age-mass}.  Of the local reference sample, only the two very massive clusters in NGC\,7252 (see position indicated by the arrows) have observed colors and luminosities comparable to the older clumps. Their location in the diagram also suggests that of the local complexes the knots in blue compact galaxies and the Antennae seem to be good progenitor analogues of these evolved clumps.

\subsection{Simulating local galaxies at the redshift of Sp\,1149}
\label{sim-highz}

From the comparison of the luminosity properties we observe that none of the local cluster complexes in our reference sample have ages (colors) comparable to the more evolved clumps of Sp\,1149. This evidence suggests that complexes in local galaxies are destroyed faster than in high $z$ galaxies. The reason for such short life--time could be related to the low masses of the complexes if compared to the clumps. 

In local galaxies, star formation is observed to happen in hierarchical regions, and, in general, giant molecular cloud (GMC) complexes show sizes up to $\sim$1 kpc \citep[e.g.][]{2011EAS....51...31E}. Studies of the clustering in local star-forming galaxies show  tight spatial and temporal correlations between massive stars and clusters at young ages \citep[e.g.][]{2007MNRAS.379.1302B}. However this hierarchical structures are rapidly erased with time. 

To test how local star-forming complexes would look at high $z$ we have simulated images of local galaxies (M\,83, Antennae, Haro\,11) to have the same depth and resolution as observed for Sp\,1149 (we remind the reader that the linear physical scale of the latter has been magnified by $\sqrt8\times$ and the fluxes by a factor of $8\times$). Following the method presented in \citet{2008ApJ...677...37O}, we have first resampled each galaxy science frame to a pixel size which matches the angular resolution of the the Sp\,1149 data (both smoothing and rebinning the frames by the corresponding quantity). As second step, we have taken into account the dimming of the surface brightness as function of the luminosity distance. The final rescaled images are included in Figure~\ref{image}. We simulated two filters for each galaxy, $U$ and $V$-band. One can clearly see that at the distance and resolution of Sp\,1149 the star-forming regions in local galaxies blend in clumps, similarly to what we observe in our high $z$ target. The brightest clumps are, however, all located in correspondence of the youngest star-forming regions in the galaxies. For example, in the case of M\,83, a comparison between the location of the bright clumps in the high $z$ $U$-band image and the contours of pronounced H$\alpha$ regions presented in Figure 3 by \citet{2012A&A...537A.145S} shows that they nicely trace each others. A similar trend is observed in the Antennae \citep[e.g. compare the location of the clumps with the color-composite image in Figure 1 by][]{2010AJ....140...75W} and in Haro\,11 \citep[e.g., Figure 1][]{2009AJ....138..923O}. 

Our simple exercise suggests that, indeed, the bright clumps in our local reference sample are very young (emitting H$\alpha$) or still star-forming and, if present at all, older clumps would clearly be less luminous. Therefore, in spite of the observational biases, the high $z$ Sp\,1149 clumps appear to be in some aspects intrinsically different. The young clumps of Sp\,1149 resemble quite closely the cluster complexes observed in local galaxies. Their difference in mass (luminosity) could be caused by the poor resolution at high $z$. The older clumps of Sp\,1149 are more massive and long-lived than any of the local complexes in our sample. The star-forming knots of blue compact galaxies are good progenitor analogues candidates of these evolved clumps. However, to test this scenario one should also be able to detect such compact knots in local post-starburst environments and to date none as been found. Finally, the two very massive clusters of NGC\,7252 have similar luminosity and color to the old clumps, showing that also at low-redshift it is still possible, under extreme environmental conditions (e.g. a galaxy merger), to form very massive systems. However, in the following section we will see  that  these clusters are structurally quite different from the Sp\,1149 clumps.

\subsection{Mass versus size relation}

In M\,51, \citet{2005A&A...443...79B} found that the stellar mass of the cluster complexes scales with their size, i.e. $M \propto R^2$, in a similar way as for the GMCs of the Milky Way and nearby galaxies \citep[][]{1987ApJ...319..730S, 2008ApJ...686..948B}. This relation, assuming that the GMCs are in virial equilibrium, suggests that they have constant surface gas density, i.e. $\Sigma \approx 100-200 \msun$pc$^{-2}$. Therefore, at least in nearby galaxies, GMCs appear to form under similar physical conditions. Interestingly, the mass--size relation does not hold for young star clusters \citep{2005A&A...431..905B}, formed in the denser regions of the GMCs. Because of turbulence acting as a scale-free process, collapsing GMCs fragment in dense clumps and cores. As such, not the single star clusters but the whole complex which has formed out of the GMC gas reservoir \citep[assuming a certain star formation efficiency, e.g.][]{2005A&A...443...79B}, can be considered as tracers. 


\begin{figure}
\includegraphics[scale=0.45]{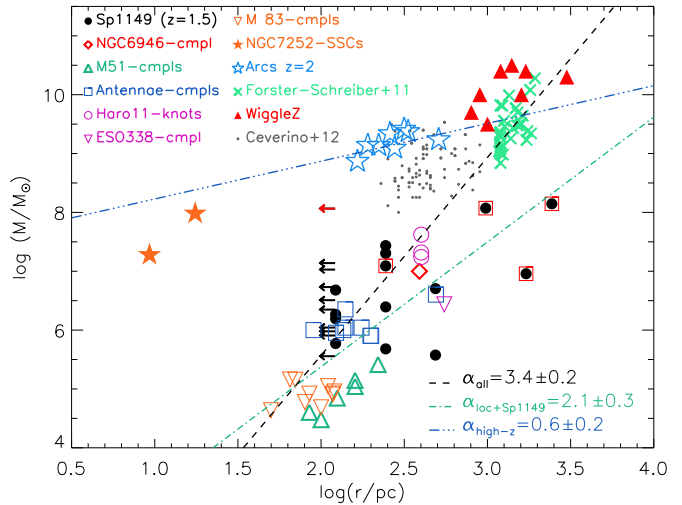}
\caption{Size versus mass relation plot. Both the high $z$ and low-$z$ samples are included and plotted with different symbols as indicated in the inset. The Sp\,1149 clumps for which we could determine a size are plotted as black dots. The dots enclosed by red squares are clumps of age $\geq 50$ Myr. The sources with only an upper limit to their sizes are shown as arrows (the red arrow shows the position of the bulge-like object) and have not been included in the fit. The SSCs in NGC\,7252 and the sample of clumps obtained from the simulations of \citet{2012MNRAS.420.3490C} have also been excluded from all the fits. The lower right insets shows the recovered slopes from the fit of different samples. The $\alpha_{\textnormal{all}}$ has been obtained fitting all the clumps and complexes, in spite of their redshift. The $\alpha_{\textnormal{loc+Sp1149}}$ is the slope of the sample including only the local reference sample and the clumps of Sp\,1149. On the other hand, the $\alpha_{\textnormal{high-z}}$ has been obtained fitting only the high $z$ sample, excluding the Sp\,1149 clumps.  \label{size-mass}}
\end{figure}


In this context, thanks to the resolution achieved in Sp\,1149, we can try to understand the physical conditions which have lead to the formation of its clumps. Using the constrained physical properties,  we attempt to build a mass--size plot, where we include also the properties of the local cluster complexes and of the high $z$ sample. The plot is shown in Figure~\ref{size-mass}. Although the whole dataset is included in the plot, we point out that the two SSCs of NGC\,7252 have been excluded from the fit because as pointed out above star clusters do not follow the mass--size relation. Similarly, the data obtained from \citet{2012MNRAS.420.3490C} have been included only for illustrative purposes. Finally, the Sp\,1149 clumps for which we derive un upper limit to their size have also been excluded from the fitting analysis. We first attempted a single fit to the selected sample. We recovered a slope much steeper than expected from the analysis of GMCs and complexes in local galaxies and from theoretical expectations. Indeed, one can see that for a range of radii between a few hundreds and a few thousands of pc, the mass of the systems span a range of 4 order of magnitude (between $10^6$ and $10^{10}$ $\msun$).  We divided, then, the sample in two subgroups. The high $z$ group with very massive knots and the local group of cluster complexes. Because of the position in the diagram, we included the Sp\,1149 clumps to the latter sample. A new fit to the two subgroups produces a power--law index of $\sim2$ for the local cluster complexes, including the clumps of Sp\,1149 and a slope of 0.6 for the high $z$ sample. Although, the scatter around the recovered slopes is large, the trend is still meaningful. The clumps of Sp\,1149 follow the mass--size relation, suggesting that their formation process is very close to the same hierarchical process which governs the formation of cluster complexes in the local universe. The high $z$ clumps from the literature sit systematically above the relation. We emphasise  that the masses of the \citet{2010MNRAS.404.1247J} and \citet{2012MNRAS.422.3339W} samples have been derived from the velocity dispersion of the H$\alpha$ emission line. Therefore, they could be overestimated if the system is not in virial equilibrium or if the line is broadened by other factors (outflows, H{\sc ii}-region shell expansions, etc.). On the other hand, the masses of the clumps of the \citet{2011ApJ...739...45F} sample, have been obtained from a combination of broadband colors and M/L ratios (similarly to our technique). However, we notice that this is, among the literature data used in this work, the most distant sample, i.e., at $z\sim2$. Either resolution effects could blend clumps to fewer and more massive, or detection limits may favour the selection of more massive host galaxies at higher redshift, with clumps forming in more extreme environments.  

In a recent work, \citet{2012arXiv1209.5741L} show evidence of a relation between the size and the SFR of the clumps (or their H$\alpha$ surface brightness). They observed a systematic offset toward higher surface brightness at higher redshift. This trend could be understood in a scenario where high $z$ galaxies have disks with higher gas surface density able to form more massive clumps. Indeed, we observe that the galactic SFRs of our high $z$ sample, which can be considered an indirect measurement of the total gas surface density of the galaxy \citep[e.g.][]{2008AJ....136.2846B}, are between a factor of 2 and 20 higher than in Sp\,1149.  Therefore, the vertical shift of the high $z$ sample in the mass--size relation could be an evidence of different conditions under which those clumps have formed \citep[e.g.][]{2012arXiv1209.1396S}. However, we stress that this offset may raise because of the different methods used to derive the masses.

\begin{figure}
\includegraphics[scale=0.45]{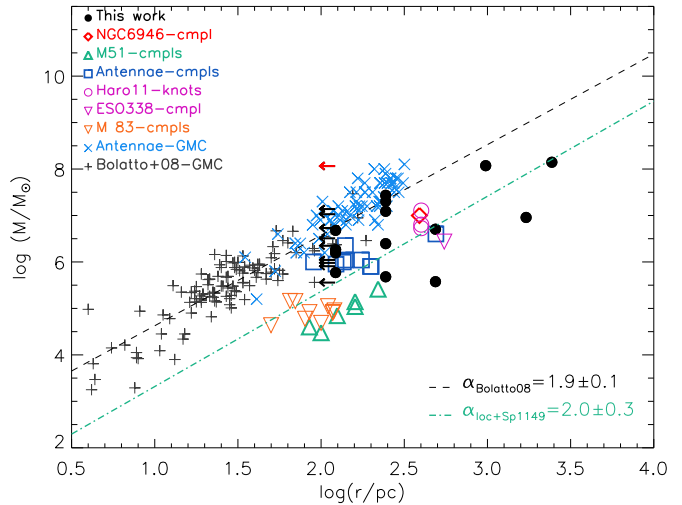}
\caption{The recovered size versus mass relation using the local and Sp\,1149 cluster complexes (as in Figure~\ref{size-mass}). The plot includes also a small sample of local GMCs from \citet{2008ApJ...686..948B} and \citet{2012ApJ...750..136W}. The fit shown as a black dashed line is obtained only from the Bolatto et al sample.   \label{size-mass-loc}}
\end{figure}

In Figure~\ref{size-mass-loc}, we compare the mass--size relation obtained for the cluster complex sample and the clumps of Sp\,1149 whit a local sample of GMCs. As already noticed by \citet{2005A&A...443...79B}, the vertical offset between the GMC sample and cluster complexes in local spirals could be due to the star formation efficiency (only a fraction of the gas in the GMCs will form stars). A considerable fraction of complexes of merger systems and of Sp\,1149 (including a large fraction of clumps with upper limits on the size) are offset with respect to complexes in local spirals. This trend could be an evidence that these more massive systems have formed in GMCs with higher gas surface density (i.e. more massive). The bulge-like clump shown as a red arrow, deviates clearly from the relation, suggesting a different mechanism for its formation as discussed in the following section.

\section{Clump formation and survival at high $z$}

As discussed in the Section 1, clumps at high-$z$ have likely formed in marginally unstable disks, in large gravitational instabilities necessary to overcome the high turbulence of the gas. In the local Universe, only merging systems, e.g. the Antennae, appear to be able to form very massive GMCs \citep[e.g.][]{2012ApJ...750..136W}, between $10^7$ and $10^8 \msun$. The formation of such massive GMCs could be related to the enormous external pressures that the merger event causes on the gas. Star-forming galaxies like the Milky Way, form GMCs via agglomeration reaching maximum masses $\lesssim10^7 \msun$ \citep{1997ApJ...476..166W}. In this context, we try to understand how the clumps in Sp\,1149 have formed. 

First of all, we focus on the young population of clumps (age $<$ 30 Myr). These clumps have formed during the recent star forming events, and are likely related to the current SFR. \citet{2009ApJ...707L.163S} reported a current SFR for Sp\,1149 of $\sim6 \msun$/yr from the $U$ band rest-frame fluxes, while using the total H$\alpha$ luminosity, \citet{2012arXiv1209.5741L} found a slightly smaller value, i.e. $\sim 1.5 \msun /yr$. Thus, the SFR of Sp\,1149 is not very different from the SFR in local spirals. Using the optical rest-frame  we derive a crude approximation of the radius of Sp\,1149, i.e. 6 kpc. Therefore, the density of SFR is $\Sigma_{SFR} \sim 0.05 \msun$/yr/kpc$^2$. Using the \citet{1998ApJ...498..541K} relation we can derive the gas surface density in the galaxy, i.e. $\Sigma_{gas}= (\Sigma_{SFR}\times A^{-1})^{1/n} \sim 50 \msun$/pc$^2$. This value is a factor of 5 larger than the mean value measured in the Milky Way ($\sim 10 \msun$/pc), but comparable to the values  observed in M\,51 and M\,83, where GMC masses can reach  a few times 10$^7 \msun$. This rough estimate shows that the young clumps we observe have mean properties in agreement with the environment of Sp\,1149, assuming star formation efficiency of 10-50 \%. It also suggests that the current star formation process in Sp\,1149 is not very different from what we observe in local spirals.

To trace the formation conditions of the more evolved clumps, with ages $> 30$ Myr, we need to consider the evolution and the mass loss that these regions have experienced.  Analytical models of local GMCs suggest that these structures are destroyed in a few tens of Myr because of stellar feedback \citep[e.g.][]{2010ApJ...709..191M} and have star formation efficiency of a few percent. Similarly, more massive high $z$ clumps should expel most of their initial gas mass because of stellar radiation and momentum-driven winds. Different conclusions are reached by \citet{2010MNRAS.406..112K}. Assuming that the star formation efficiency per free-fall time is of a few percent, then clumps can retain a large fraction of their initial gas and convert it into stars. They also suggest that high $z$ clumps may lose $\sim$ 50\% of their initial stellar mass because of tidal stripping during their migration toward the centre. We use the \citet{2010MNRAS.406..112K}'s model to try to derive an approximate initial mass of the older clumps in Sp\,1149. We divide the clumps older than 30 Myr in 3 groups; 1) 30$<$age$<$50 Myr, in this bin the minimum and maximum clump mass is $5\times10^6 \msun$ and $3\times10^7 \msun$, respectively (the bulge has been excluded); 2) 50$\leq$age$<$100 Myr, clump mass of $\sim 1\times 10^7 \msun$; 3) age $\geq 100$ Myr, clump mass between $2\times 10^7$ and $1\times10^8 \msun$. Assuming that 20\% of the formed stellar mass is lost because of stellar evolution and 50 \% because of the interaction with the tidal field of the galaxy, the initial stellar clump mass should be a factor of $\sim$ 2.4 more massive than currently observed. According to the \citet{2010MNRAS.406..112K}'s model the massive clumps should be able to convert most of their initial gas into stars, under the condition that the star formation efficiency per free-fall time is not high. Therefore, we expect that the initial GMCs from which massive clumps have formed should have been at least a factor of $\sim$3 more massive than the current observed mass.

\begin{figure*}
\includegraphics[scale=0.6]{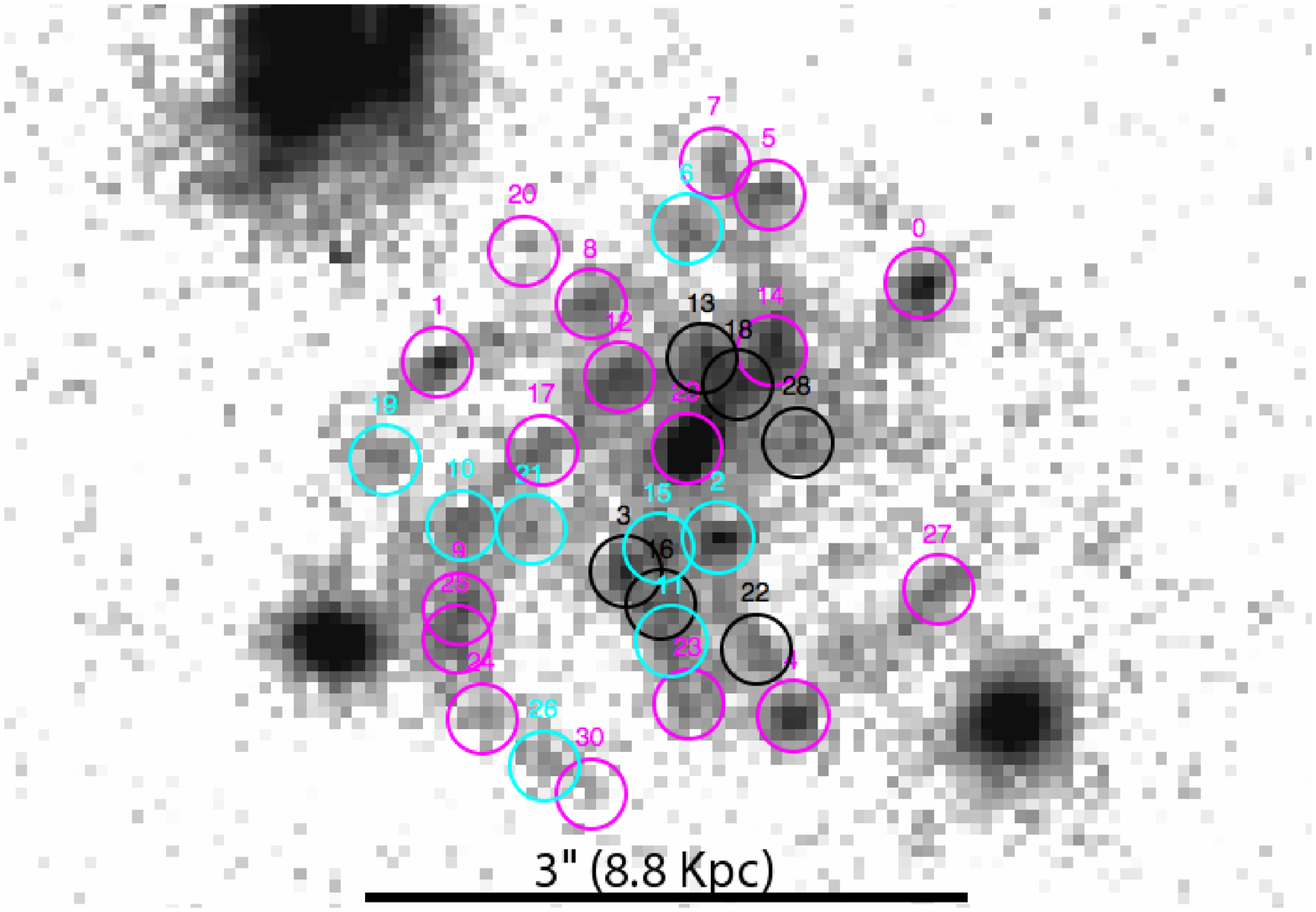}\\
\includegraphics[scale=0.43]{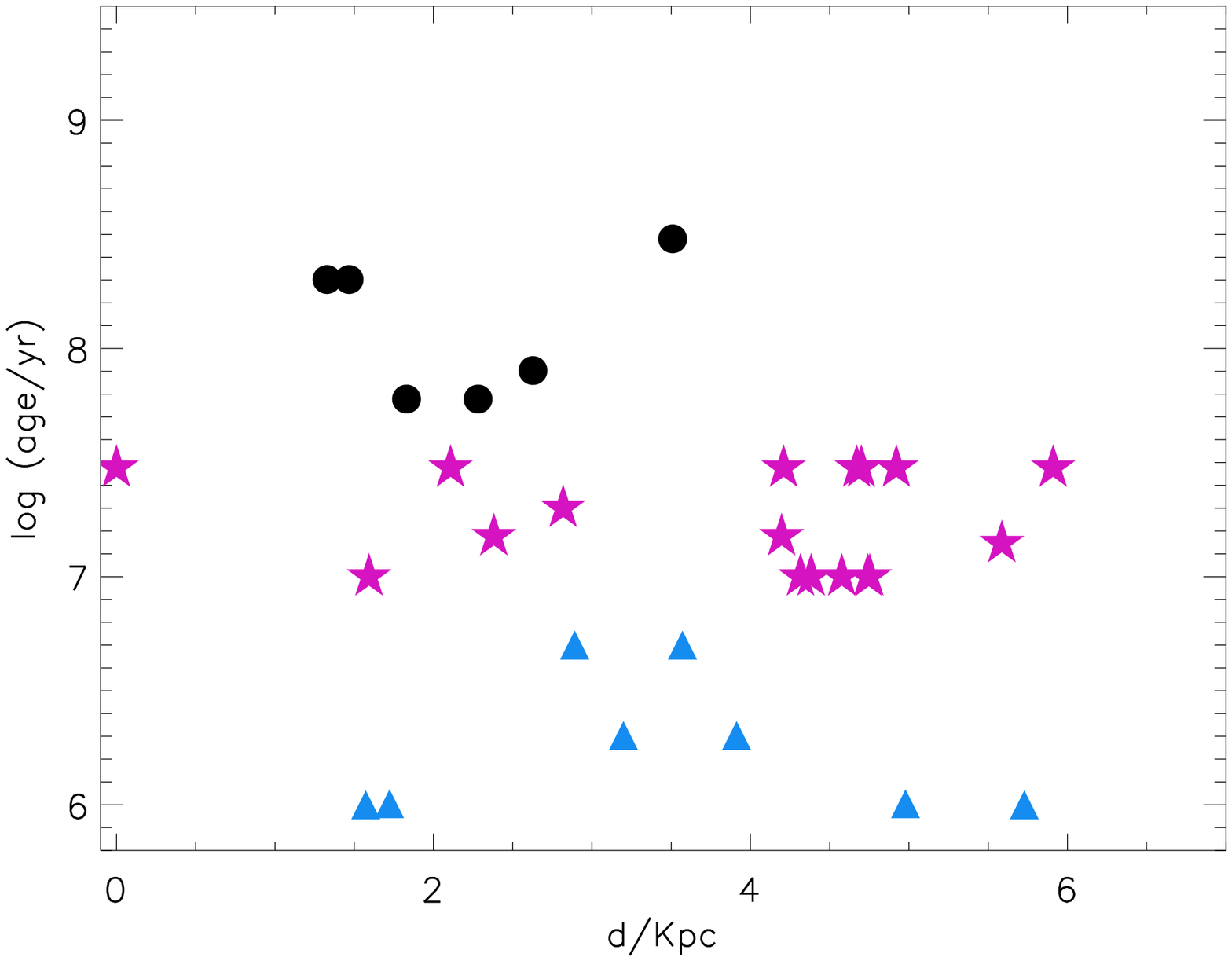}
\includegraphics[scale=0.43]{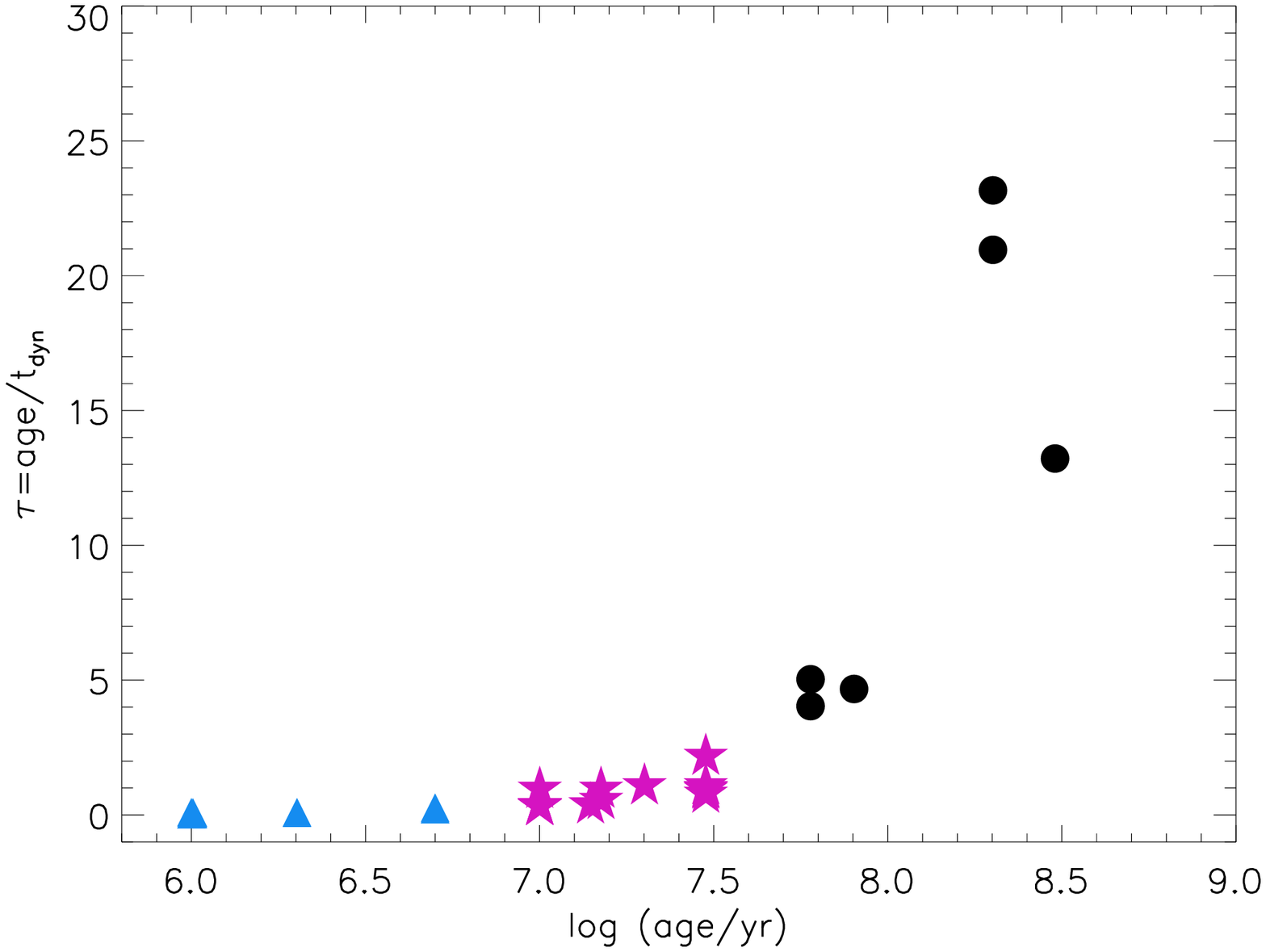}
\caption{Top panel; Clumps detected in Sp\,1149. The clumps have been divided in 3 age ranges, shown with different symbols and colors; cyan (triangles in the bottom plots) for ages $<$ 30 Myr; purple (stars) for clumps with ages between 30--50 Myr; black (filled dots) are systems older than 50 Myr. The size of the circles correspond to the aperture radius used to perform photometry. Bottom left panel; Clump age versus projected distance from the centre of the galaxy. Bottom right panel; Total dynamical time for each clump ($\tau = $age/t$_dyn$) plotted as function of their ages. The color-coding is the same as above. \label{pos}}
\end{figure*}

If we correct the current stellar mass by the 2.4 factor to take into account the stellar mass loss and assume that the mass has formed in about 10 Myr (this choice is consistent with the stellar evolutionary models used to derive the stellar masses), we can derive a rough estimate of the SFR in these more massive  and evolved clumps. We obtain that for ages between 30 and 50 Myr the averaged clump SFR $\sim 3 \msun$/yr; in the range 50 to 100 Myr $\sim 3 \msun$/yr; and $\sim 22.0 \msun$/yr in clumps with ages older than 100 Myr. If we consider that each clump contributes to a fraction of the total SFR in the host system, then this increasing value of the SFR as function of time may indicate a change in the conditions under which clumps have formed in the past (higher surface gas density, larger gravitational instabilities) and are forming (see above, e.g. similar conditions to local star-forming galaxies) in Sp\,1149. However, we stress that this is only a suggestion and that it depends of the many assumptions we have made on deriving the clump properties and the initial stellar masses.

\subsection{Bulge formation in Sp\,1149}

In high $z$ galaxies, the formation of the bulge seems to be a gradual process. \citet{2009ApJ...692...12E} observed that a considerable fraction of clump clusters and chains have red and massive clumps which could be considered bulge-progenitors. These young bulges are a few times more massive than the other star forming clumps in the system but the ages are similar. In young spirals, bulges are older and about 20--30 times more massive than the clump population, suggesting a clear evolution. Simulations \citep[e.g.][]{2008A&A...486..741B, 2010MNRAS.404.2151C} show that clumps migrate toward the centre of the system because of dynamical friction and tidal torques in about 10 dynamical times, which corresponds to a timescale of about 0.5 to 1 Gyr. There, clumps  coalesce and slowly build up the bulge and the thick disk component \citep{2007ApJ...670..237B, 2008ApJ...688...67E}. The migration of clumps should cause a significant age spread, with older clumps preferentially closer to the centre. Indeed, evidence of such trend has already been reported in previous clump studies \citep[e.g][]{2009ApJ...692...12E, 2011ApJ...739...45F}. 

In this context, we try to use the recovered clump properties of Sp\,1149 to test the bulge formation scenario. Source \# 29 is most likely the young bulge of Sp\,1149. Already \citet{2009ApJ...707L.163S} observed that the bulge does not dominate the luminosity of the galaxy, suggesting its early formation stage. The fit to its SED gives a best age estimate of 30 Myr and a mass of a few $10^8 \msun$, in good agreement with the age and mass range reported by \citet{2009ApJ...692...12E} for young bulges in spirals at $z\sim1.5$. The age and mass determined by using the models with 100 Myr continuous SFR do not change significantly (age of 40$^{+0}_{-39}$ Myr, mass of 6$^{+1.7}_{-5.1} \times 10^{7} \msun$)  our conclusions. The fact that the progenitor bulge is younger than other clumps in the galaxy is not in contrast with the scenario of bulge formation presented above. Indeed, star formation can proceed for a longer time in the central region than in single clumps without preventing the accretion of older systems \citep{2011EAS....51...59E}. The youth of the bulge is also supported by narrow band H$\alpha$ imaging and spectroscopy of the galaxy, where it appears to be quite a bright H$\alpha$ source \citep{2012arXiv1209.5741L, 2011ApJ...732L..14Y}. 

In Figure~\ref{pos}, we show the position of the detected clumps in the galaxy (top panel); the projected distances (bottom left) and the total dynamical time (bottom right) that the clumps have survived ($\tau = $age/t$_dyn$) in the galaxy as function of their age. 
The oldest clumps (age$>$50 Myr) are all at a distance less than 4 kpc. Defining the dynamical time as the ratio between the rotation velocity and the radius, $t_{dyn} \propto r/v_{rot}$, we can get an rough estimate of the possible migration of the clumps \citep[e.g.][]{2011ApJ...739...45F}. \citet{2011ApJ...732L..14Y} found a rotation velocity for Sp\,1149 of $\sim 150$ km/s. For the two old clumps at 1.5 kpc distance, we estimate a $t_{dyn}\sim 10$ Myr. This implies that they have survived about 20 $t_{dyn}$, and most likely have migrate to their current position closer to the centre. For the remaining old clumps, we see that they have ages about 5--15 times $t_{dyn}$, still supporting the scenario of long-lived migrating clumps.  The remaining younger clumps have ages of maximum a few $t_{dyn}$, suggesting that their current position is close to where they formed, as expected for young cluster complexes in nearby star-forming galaxies. We notice that 4 old clumps out of 6 for which we have size measurements are among the largest systems (see black dots encompassed by red squares in Figure~\ref{size-mass}), suggesting possibly an ongoing dissolution of these systems while approaching the centre.

\subsection{Formation of globular cluster progenitors}

In a recent work, \citet{2010MNRAS.403L..36S} suggested that star clusters forming in clumps at $z\sim2$ may form the metal-rich globular cluster population which is usually associated with the thick disk and bulge component of late spiral galaxies. We closely follow their method, adding, however, some extra information on the cluster formation efficiency of each clump, i.e. the fraction of star formation happening in clusters per clump. There is observational evidence that this fraction \citep[referred to as $\Gamma$,][]{2008MNRAS.390..759B}  of clusters that a galaxy may form scales with the global SFR of the galaxy  \citep[e.g][]{2010MNRAS.405..857G, 2011MNRAS.417.1904A}. Analytical models suggest that such trend is naturally reproduced if the hierarchical mass spectrum of the ISM is combined to the disk stability  ($Q$-parameter) and the limit imposed by the environment on the dissolution of clusters \citep{2012arXiv1208.2963K}. For the younger clumps (age$<$ 30 Myr), we use the current galactic $\Sigma_{SFR}$ and the relation produced by \citet{2010MNRAS.405..857G} to derive $\Gamma \approx 14$ \%. Therefore, we will assume that only 14 \% of the estimated stellar mass of the clumps will form bound clusters.  For the older clumps we don't have direct measurements of the galactic SFR at the moment they formed, but we have seen in the previous section that blue compact galaxy knots could be possible progenitor analogues  of these clumps. \citet{2011MNRAS.417.1904A} found $\Gamma \approx 40-50$ \% in these systems, so we assume that 50 \% of the initial total stellar mass is in globular cluster progenitors. We approximate the cluster mass function to be a power-law of the form, $dN \propto M^{-2}dM$. The minimum and maximum cluster mass that a clump can form are fixed to $10^4$ and $(M_{clump}\times \Gamma)$, respectively. The latter assumption means that the maximum mass of a cluster can be as large as the total fraction of stellar mass of the clump going into clusters. After integrating, we obtain that $\sim 300$ star clusters with masses larger than $10^4 \msun$ have been formed in the young clumps. In the older clumps (age$>$ 30 Myr) we estimate that about 31000 young star clusters have been produced. However, as discussed in \citet{2010MNRAS.403L..36S}, many of the low mass clusters will dissolve because of two-body relaxation, while the high-mass end of the power-law is better described by a Schechter function, with a characteristic high-mass cut off. Taking these factors into account they estimate that for every 7000 star clusters only a small fraction ($\sim 120$) will survive until  $z\approx0$ as globular clusters. Therefore, we predict a lower limit number of about 540 metal-rich globular clusters in Sp\,1149 in good agreement with expectation (Shapiro et al. predicted a metal-rich globular cluster population up to $\sim$ 600 systems for $z\sim0$ galaxies with stellar mass of a few times $10^{11} \msun$).

\section{Conclusions}

Our single-object study has revealed some useful insights into the intermediate evolutionary stage in which clumpy turbulent systems become modern spiral galaxies.

Sp\,1149 is not a special galaxy. Its current SFR (1--6 $\msun$/yr) is similar to what is observed in local spirals and the clumpy appearance is not much different than in other spirals observed at similar redshift. What makes Sp\,1149 an interesting case study is its position behind a massive galaxy cluster. The magnification gained  by the gravitational lens has allowed us to study the properties of the clumps at a spatial resolution $\sim$3 times higher than other systems at a similar redshift. Thanks to the achieved resolution, we have analysed what we believe to be the transient evolutionary phase that high $z$, intense star-forming galaxies experience when they evolve into more quiescent massive systems, as we observe in the local Universe. 

Using the CLASH survey, we detected roughly 30 clumps in Sp\,1149, for which we derived ages, masses, extinctions, and effective radii. To allow a comparison with local star-forming galaxies, we collected available literature data of cluster complexes with properties of star-forming knots in luminous blue compact galaxies and some of the most massive studied intermediate age SSCs.

We found that the clumps in Sp\,1149 are typically more luminous than cluster complexes in nearby spirals, similar to the cluster complex luminosity of the Antennae, but not as bright as star-forming knots in blue compact galaxies. Curiously enough the median rest-frame $U$ and $V$ band magnitudes of the clumps are quite similar to the massive cluster complex in the nearby spiral NGC\,6946, which gives a close idea of the morphological aspect and physical properties of the young clumps. Diagnostic tools like the color-magnitude and the age--mass diagrams show the coexistence of  two populations of clumps in Sp\,1149, with an age-spread up to a few hundreds of Myr.

The older clumps are the most massive and luminous. None of the nearby cluster complexes has color (ages) comparable to this subsample of old clumps. This suggests that cluster complexes dissolve faster, in contrast to older clumps, which have formed in more extreme environmental conditions and have masses (properties) that ensure their survival for a significant amount of galactic dynamical timescales. If we extrapolate the luminosity of these older clumps at the moment they formed they must have been at least 2 magnitudes brighter than currently observed. We suggest that among the nearby reference sample, knots in blue compact galaxies could be considered progenitor analogues  of these evolved population. 

The younger clumps (age $<$ 30 Myr) have colors which overlap quite well with the local sample of cluster complexes, supporting their young age estimate. The masses are a factor of 10 larger than typical cluster complexes in local spirals and comparable to complexes in the Antennae and NGC\,6946. In our imaging simulation of local galaxies at the distance and resolution of Sp\,1149, we observed that indeed the cluster complexes are all distributed along the bright H$\alpha$ regions, i.e. they have recently formed. The morphological detail of the single complexes are erased, and we don't see any major difference between them and the young clumps in Sp\,1149, suggesting that the difference in mass and luminosity could be caused by the limited resolution we can achieve in the latter system. 

We tried to understand how clumps have formed, looking at the mass-size relation, originally constrained on GMCs of the Milky Way by \citet{1987ApJ...319..730S}, but similarly reproduced by cluster complexes in M\,51 \citep{2005A&A...443...79B}. In its original formulation this relation suggests that GMCs have similar gas surface densities. Cluster complexes still follow the relation because they are scaled down (in total mass) descendants of the GMCs. We found evidence that the clumps of Sp\,1149, together with the complex reference sample at $z\sim0$, follow the trend of the mass-size relation.

Given the current $\Sigma_{SFR}$ of the galaxy and using the Kennicutt  law we derive a galactic surface gas density of $\sim$50 $\msun$/pc$^{-2}$, similar to M\,83 and M\,51, where GMCs can reach masses of $\sim10^6-10^7 \msun$. These are most likely the GMC progenitors from which the young clumps of Sp\,1149 have formed, assuming star formation efficiency of 10-50 \%. However, we find evidence that the older clumps have formed in more massive GMCs. We tried to reconstruct the initial conditions for these evolved systems, assuming that they have lost 50 \% of their initial stellar mass because of tidal interaction with the diffuse stellar population of the disk and about 20 \% of their initial stellar mass because of stellar evolution. We obtained that the GMCs from which these clumps formed should have been at least a factor of 2.4 more massive than their current derived properties. We derived that the SFR in the old clumps, at the moment they formed, was in the range $3-22 \msun$/yr, suggesting that the total SFR in the galaxy should have been significantly higher than what is measured currently. The galaxy is experiencing a slow decline in SFR and a likely transitional phase toward more quiescent star-formation modes.
 
The presence of long-lived clumps allowed us to test another important element in the evolution of young spirals into a more evolved system, i.e. the formation of the bulge. Sp\,1149 has already a young bulge, suggesting that the formation process is ongoing. We found that clumps older than $50$ Myr are all located at a projected distance smaller than 4 kpc and have survived between 6 and 20 galactic dynamical times. Their current location in the galaxy must be quite different from their birthplace, producing evidence of an inward migration toward the centre, where they are expected to build up the bulge. On the other hand, the dynamical timescales of the younger clumps are significantly shorter, meaning that they are still or quite close to their birthplace, i.e. along the spiral arms, similar to the cluster complexes observed in nearby systems.

Finally, we attempted to estimate whether the clumps we observe in Sp\,1149 could produce the expected metal-rich globular cluster population usually associated with the bulge and thick disk components of local spirals. We found a consistent fraction of star clusters which will survive and eventually evolve into globular clusters. We notice that the large majority of clusters which will survive are produced in the long-lived, more massive clumps. A large fraction of clusters which are formed in complexes younger than 30 Myr most likely will not survive long enough. The difference could be understood in terms of size-of-sample effects. More massive clumps have higher probability than cluster complexes to sample the cluster mass function at the high mass bins. These more massive clusters have higher probability to survive up to an Hubble time.

\acknowledgments
We thank the referee for insightful comments to the manuscript. A.A. thanks A. van Der Wel, E. Lusso, and S. Meidt, for useful discussions and hints on the analysis. S.S. Larsen is acknowledged for the access to the $U$ and $V$ band FORS-2 data of M\,83. T. Jones is thanked for providing the catalogue of clumps in lensed galaxies at redshift $\sim2$. G.\"O. is a Royal Swedish Academy of Sciences Research Fellow, supported by a grant from the Knut and Alice Wallenberg foundation. G.\"O. and E.Z. acknowledge financial support from the Swedish Research Council and the Swedish National Space Board. RCL acknowledges a
studentship from STFC.



{\it Facilities:}  \facility{HST (ACS, WFC3)}




\clearpage




\end{document}